\documentclass{iopart}
\usepackage{graphicx}
\usepackage{amsfonts}
\usepackage{amssymb}
\usepackage[symbol*]{footmisc}
\newcommand {\be} {\begin{equation}}
\newcommand {\ee} {\end{equation}}
\newcommand {\Be}{\begin{eqnarray*}}
\newcommand {\Ee} {\end{eqnarray*}}
\newcommand {\bey} {\begin{eqnarray}}
\newcommand {\eey} {\end{eqnarray}}
\newcommand{\bit}{\begin{itemize}}      
\newcommand{\eit}{\end{itemize}}
\newcommand{\bfl}{\begin{flusleft}}
\newcommand{\efl}{\end{flusleft}}
\newcommand{\bfr}{\begin{flushright}}
\newcommand{\ec}{\end{center}}
\newcommand{\ben}{\begin{enumerate}}    
\newcommand{\een}{\end{enumerate}}

\newcommand{\bx}{{\bf{x}}}

\newcommand{\bof}{{\bf{f}}}
\newcommand{\R}{\hbox{I \kern -.5em R}}

\newcommand{\mtM}{\mathcal{M}}
\newcommand{\mtT}{\mathcal{T}}
\newcommand{\mtD}{\mathcal{D}}

\begin{document}


\title{Covariant Lyapunov vectors}

\author{Francesco Ginelli}
\address{SUPA, Institute for Complex Systems and Mathematical Biology, King's College, University of Aberdeen, Aberdeen AB24 3UE, United Kingdom}
\author{Hugues Chat\'e}
\address{Service de Physique de l'\'Etat Condens\'e,
  CEA-Saclay,~91191~Gif-sur-Yvette, France}%
\author{Roberto Livi}
\address{Dipartimento di Fisica, INFN and CSDC, Universita` di
  Firenze, via G. Sansone 1, 50019 Sesto Fiorentino, Italy and
  ISC-CNR, via Madonna del Piano 10, I-50019 Sesto Fiorentino, Italy}
\author{Antonio Politi}
\address{SUPA, Institute for Complex Systems and Mathematical Biology, King's College, University of Aberdeen, Aberdeen AB24 3UE, United Kingdom}

\date{\today}

\begin{abstract}
The recent years have witnessed a growing interest for covariant
Lyapunov vectors (CLVs) which span local intrinsic directions in the phase
space of chaotic systems. Here we review the basic results of ergodic theory, 
with a specific reference to the implications of Oseledets' theorem 
for the properties of the CLVs. We then present a detailed description
of a ``dynamical'' algorithm to compute the CLVs and show that it generically
converges exponentially in time. We also discuss its numerical performance and compare
it with other algorithms presented in literature. We finally illustrate how
CLVs can be used to quantify deviations from hyperbolicity with reference to a 
dissipative system (a chain of H\'enon maps) and a Hamiltonian model (a
Fermi-Pasta-Ulam chain).
\end{abstract}
\maketitle
\section{Introduction}

Lyapunov exponents (LEs) have been recognized as a central tool for the 
characterization of chaotic dynamics since the late '70s when effective
algorithms have been independently proposed by Shimada and Nagashima \cite{Shimada}
and Benettin et al. \cite{Benettin}. In fact, although it had been theoretically
clarified that LEs provide the right framework for an appropriate 
(coordinate-independent) characterization of dynamical regimes (see the
pioneering work by Oseledets \cite{osel}), it was necessary to wait 
until an effective algorithm was made available to a wide community of 
scientists, to appreciate the usefulness of such a tool in physically
relevant models, where it is not obvious whether theoretical results
apply. It is nowadays widely recognized that LEs help to
quantify a number of interesting physical properties such as dynamical
entropies and fractal dimensions \cite{EckmannRuelle}.

Since the LEs can be viewed as the eigenvalues of suitable matrices,
one might have expected a comparable interest to be devoted
to the corresponding Lyapunov vectors. Unfortunately, for a long
time, the only numerically accessible vectors were those ones 
obtained as a by-product of the Gram-Schmidt orthogonalization
procedure (an intrinsic step of the algorithms defined in \cite{Shimada,Benettin}),
which are coordinate-dependent. As a result, Lyapunov vectors did
not attract much interest within the nonlinear-dynamics community,
even though an objective definition of intrinsic,
covariant Lyapunov vectors\footnote{Sometimes they are also refereed
  as {\it characteristic} Lyapunov vectors.} (CLVs) has been given 
three decades ago \cite{ruelle}. In fact, until recently, as a result
of the lack of effective algorithms, only a few papers can be found
in the literature where the proper CLVs have been determined
and used to characterize chaotic states \cite{Abarbanel, Legras, Politi98}.

The situation has drastically changed a few years ago, when efficient
algorithms have been developed \cite{Ginelli2007, WS, KP}, so that 
many scientists are now aware that CLVs can offer information on the 
local geometric structure of chaotic attractors, as
opposed to LEs, which are powerful but global quantities. 
Several papers appeared, where CLVs have been successfully 
employed to better understand many aspects of chaotic dynamics 
\cite{Pazo, KK, Radons08, Takeuchi2009, Takeuchi1, Posch, Morriss11,
Ouchi11, Koba12, Viana2012, Radons2012}. Some of the relevant questions are touched
in this Special Issue.
 
The goal of this paper is to provide a fairly general introduction
to the topic, including a description of the various algorithms
that have been proposed, but mostly focusing on the ``dynamical" approach
that we believe to be the most effective one \cite{Ginelli2007}.
In particular, we discuss the convergence properties of the CLVs
by making use of a suitable large-deviation function.
An additional point that we explore is the minimal angle between
the unstable and stable manifold, as it provides direct information on
the hyperbolicity of the underlying dynamics. In particular, we
find that angle distribution sastisfies a natural scaling law,
both in a dissipative and Hamiltonian context.

More precisely, we start with a discussion of the theoretical context where CLVs
can be properly defined. This is done in Section~\ref{sec2}, which
contains a short review of the basic results of ergodic theory in the 
generic context of smooth Riemannian manifolds.
In Section~\ref{sec3}, we then review some implications of
Oseledets' theorem in the more practical context of $\R^N$ and
discrete-time dynamics (to which every continuous-time dynamics has to be
be reduced in computer simulations) for time-reversible systems.
In Section~\ref{Sec4}, we introduce covariant Lyapunov vectors, describe 
in detail the algorithm introduced in \cite{Ginelli2007}, and discuss 
its application to non-invertible dynamics as well.
The convergence properties of the algorithm are discussed in Section~\ref{sec5}, 
with reference to coupled H\'enon maps and Fermi-Pasta-Ulam chains.
In Section~\ref{Sec6} we briefly review other algorithms that have
been proposed in the literature and mutually compare them.
Section~\ref{sec7} is devoted to a discussion of the angles between CLVs,
essentially focussing on the angle between the stable and unstable manifold
in a dissipative (coupled H\'enon maps) and Hamiltonian (Fermi-Pasta-Ulam chain)
model.
We conclude by briefly mentioning other relevant applications and pending questions.

\section{Basic tools from ergodic theory}
\label{sec2}
The mathematical foundations of ergodic theory provide a solid basis for the study of
dynamical systems and, in particular, of their stability properties. For what concerns 
the application to various fields like statistical and celestial mechanics,  plasma and accelerator physics, fluid dynamics, meteorology
etc.,   the main
contribution of ergodic theory stems from the possibility of translating rigorous mathematical
results into effective algorithms, that allow for a quantitative analysis. It is worth to point out that
the tools of ergodic theory apply to both continuous (in time) as well as to
discrete (in time)  dynamical systems, i.e. flows and maps, respectively.  The study of the
Lyapunov characteristic exponents, or simply Lyapunov exponents, 
is one of the main chapters of ergodic theory and goes back 
to the beginning of  the XX-th century, when the Russian mathematician  Aleksandr M. Lyapunov
introduced them as indicators of the stability of a singular point or of a periodic
orbit. As mentioned in the introduction, here we aim at providing 
an overview about covariant Lyapunov vectors and their practical use in applications.
Accordingly, in this section we summarize the main results of ergodic theory that are
strictly related to the concept of  CLV and to the possibility of explicitly computing them
\footnote{For a more general survey about ergodic theory we suggest the contribution
by Lai-Sang Young in this special issue}.\\

\subsection{Dynamical systems, stability  and measures}
Oseledets' theorem is a major achievement in the study of the stability of dynamical systems. It proves  the
existence of LE for generic orbits under quite general conditions \cite{osel}. In practice, this theorem 
offers the possibility of extending Lyapunov stability analysis from fixed points and periodic orbits, to 
any trajectory of a dynamical system $(\mtM, L)$ , either continuous or discrete in time, defined 
on a Riemannian manifold $\mtM$ of dimension $N$ and
equipped with a suitable metric. The application $L$ is the evolution operator of the dynamical
system, i.e.  $L^t(x) \equiv x_t$, where $x_t$ is the image at time $t$ of the initial condition $x$ at time $t=0$.
We assume that at each point $x\in \mtM$ we can identify a tangent
space $\mtT_x \mtM$. With reference to a vector $\xi \in \mtT_x \mtM$ we can define the expansion rate
\be
\gamma(\xi,x,t) = \frac{\| D L^t(x) \xi \|}{\| \xi \|}
\ee
where the symbol $\| \bullet \|$ is the norm and $D L^t(x): \mtT_x\mtM \to \mtT_{L^t(x)} \mtM$ indicates the evolution in tangent
space. The Lyapunov characteristic exponent associated to the vector $\xi \in \mtT_x\mtM$ is defined as
\be
\label{eqLy}
\lambda(\xi,x) = \lim_{t\to\infty} \frac{1}{t} \ln  \frac{\| D L^t(x) \xi \|}{\| \xi \|}
\ee
provided this limit exists. This quantity essentially tells us how an
infinitesimal perturbation $\delta x$ of the initial condition
$x\in \mtM$ along the vector $\xi \in \mtT_x \mtM$ is exponentially
expanded (or contracted) at large times
\be
\|\delta x_t \| \sim \|\delta x \| e^{\lambda (\xi,x)\,t}\,.
\ee 
The crucial point here is that for a generic (non periodic) trajectory in $\mtM$ at any $t$ the tangent
application $D L^t(x)$ applies to different tangent spaces at
different times, and the very notion of eigenvalues and
eigenvectors of this application is undefined, even if the limit
(\ref{eqLy}) exists. This notwithstanding, a filtration,
i.e. a local decomposition into suitable subspaces of $\mtT_x\mtM$ at any $x\in\mtM$, still exists as in the
case of fixed points and periodic orbits.
In fact, when the vector $\xi$ is varied in $\mtT_x\mtM$ the quantity $\lambda(\xi,x)$
takes a finite number $m \le N$ of distinct values $\lambda_1^*(x) > \lambda_2^*(x) > \cdots > \lambda_m^*(x)$.
The filtration of the tangent space into subspaces $S_i$, $\mtT_x\mtM \equiv  S_1 \supset S_2 \cdots \supset S_m$,
is such that by choosing $\xi \in S_i \setminus S_{i+1}$, one has $\lambda(\xi,x) = \lambda_i^*(x)$. It is important to
point out that in general one can associate to each $\lambda_i^*$ a multiplicity (or degeneracy) 
$g_i = \mathrm{dim} S_i - \mathrm{dim} S_{i+1}$.  A basis of $\mtT_x\mtM$, obtained by taking in
each subspace $S_i \setminus S_{i+1}$ a number $g_i$ of independent
vectors is called a {\it normal basis}. It can be easily argued that 
if  $(f_1,f_2,\cdots,f_N)$ is
a generic basis of the tangent space $\mtT_x\mtM$ and  $(e_1,e_2,\cdots,e_N)$ is
a normal basis one has
\be
\label{lambdap}
\sum_{i=1}^N \lambda(e_i,x) \le \sum_{i=1}^N \lambda(f_i,x)
\ee
where the equality holds only if $(f_1,f_2,\cdots,f_N)$ is a normal basis.
The ordered sequence of the $m$ characteristic exponents $\lambda_i^*(x)$, each one
repeated with its multiplicity $g_i$ is called the spectrum, $Sp(x)$, of characteristic Lyapunov
exponents  $\lambda_1(x) \ge \lambda_2(x) \ge \cdots \ge \lambda_N(x)$. In practice, the
existence of a filtration amounts to establishing the existence of linear subspaces 
$E_i = S_i \setminus S_{i+1}$ of dimension $g_i$ that identify the characteristic Lyapunov
exponent $\lambda_i(x)$ together with its multiplicity. One can easily realize that the extension
of such a concept to linear subspaces $E \subset \mtT_x\mtM$ with $\mathrm{dim} E = p \le N$ is
straightforward. In fact one can define the Lyapunov characteristic exponent of order $p$
\be
\label{eqLyp}
\lambda(E,x) = \lim_{t\to\infty} \frac{1}{t} \ln  \frac{\mathrm{Vol}^p ( D L^t(x) \xi_1, \cdots,D L^t(x) \xi_p) }
{\mathrm{Vol}^p( \xi_1,\cdots,\xi_p )}
\ee
provided this limit exists,
where $\mathrm{Vol}^p (\cdots)$ is the volume of the $p$-dimensional parallelepiped generated by
the tangent vectors in its argument. Let us stress that  the LE defined in equations (\ref{eqLy}) and 
(\ref{eqLyp}) are independent of the chosen metric.\\
After all of these preliminary considerations one can summarize the very content of Oseledets' theorem
that applies to a dynamical system $(\mtM,\mu,L)$ that preserves the measure $\mu$ and can be
either continuous or discrete in time, invertible or non-invertible. The theorem states that \cite{osel}

\begin{quote}
{\it For almost all $x\in \mtM$ and for any subspace $E \subset \mtT_x \mtM$, such that $\mathrm{dim} E = p \le N$
the limit in Eq.~(\ref{eqLyp}) exists and is finite  and, in particular,  the limit in Eq.~(\ref{eqLy}) exists and
is finite for any tangent vector $\xi \in \mtT_x \mtM$. Moreover the spectrum is a measurable function
of $x$ and for any subspace $E \subset \mtT_x \mtM$ it exists a normal
basis $(e_1,e_2,\cdots,e_N)$ such that}
\be
\lambda^{(p)} (E,x) =  \sum_{i=1}^p \lambda(e_i,x)\,.
\ee
\end{quote}
Some interesting consequences of this theorem are worth to be listed. First of all, one can easily
realize that LE are constant of the motion generated by the evolution operator $L$, thus implying that
for an ergodic system they are almost--everywhere constant, $\lambda_i(x)=\lambda_i$.  
If the dynamical system is invertible and
the measure $\mu$ is conserved, the sum of the Lyapunov characteristics
exponents (or the Lyapunov exponent of order $N$) is zero, $\lambda^{(N)}( \mtT_x \mtM,x)
=  \sum_{i=1}^N \lambda(e_i,x) = 0$. In particular, for a Hamiltonian system and, more generally, for any
symplectic diffeomorphism (i.e., canonical transformation of the manifold $\mtM$ onto itself) the spectrum of LE is
symmetric, i.e. $\lambda_i(x) = - \lambda_{N-i+1} (x)$ for
$i=1,\ldots, N/2$, where each individual exponent $\lambda_{N/2-j}(x) = 0$
for $ j=0,\cdots, n_c-1$ if there are $n_c$ constants of the motion (one of which is necessarily
the energy of the system). Moreover, for any continuous time dynamical
system whose support does not reduce to a fixed point, at least one of
the Lyapunov exponents should vanish.\\
Finally, it is worth pointing out one further major consequence of Oseledets' theorem that is of primary 
importance for the aim of this manuscript. We make reference, for simplicity, to
the case of discrete--time evolution, where the definition of the adjoint of the evolution 
operator in tangent space is straightforward\footnote{Notice {\it en passant} that this is the
general situation one has to deal with when numerical estimates of $Sp(x)$ and of the
corresponding eigenvectors have to be performed. In fact, even continuous--time
dynamical systems have to be integrated by transforming them into discrete-time
algorithms, whose reliability is primarily related to the conservation of a suitable
measure $\mu$ and its symmetries and to the choice of appropriate integration time
steps that guarantee a sufficient sampling all over the manifold $\mtM$ }: 
given $D L^t(x): \mtT_x\mtM \to \mtT_{L^t(x)} \mtM$ its adjoint operator 
$(D L^t(x))^*: \mtT_{L^t(x)} \mtM \to \mtT_x\mtM $ represents the time-reversed
evolution in tangent space of a reversible dynamical system $(\mtM,\mu,L)$.
By definition the operator $(D L^t(x))^* D L^t(x)$ is a symmetric, positive-definite
linear operator on $\mtT_x\mtM$ and, with the same hypotheses of Oseledets' theorem,
one can prove that the following limit exists almost everywhere
\be
\label{symts}
\mtD (x) = \lim_{t \to \infty}  [(D L^t(x))^* D L^t(x)]^{\frac{1}{2t}}\,.
\ee
Moreover, its eigenvectors $\{d_1(x),\cdots, d_N(x)\}$ are a normal basis in
$x$, while the corresponding eigenvalues $\{\delta_1(x), \cdots , \delta_N(x)\}$  are such that
$\ln \delta_i(x) = \lambda_i(x)$. Notice that the operator (\ref{symts}) describes
the evolution in tangent space forward in time for a duration $t$ and, then,
backward in time for the same lapse of time. Accordingly, the initial and
the final tangent spaces coincide, so that the difficulties inherent the extension
of the Lyapunov stability analysis to generic trajectories in $\mtM$ are removed.
Notice also that, while in ergodic systems the eigenvalues of $\mtD (x)$ are almost everywhere constant,
the corresponding eigenvectors do depend on $x$.\\

\subsection{Dynamical systems and entropies}
For the sake of completeness it is worth discussing shortly
the crucial contributions by Ya. B. Pesin to the mathematical theory of LE.  In a paper
of 1976 \cite{pesin1} he generalized the concept of Lyapunov characteristic exponents for
a family of mappings that satisfy more general conditions than Lyapunov regularity.
In practice, the main result is the identification of invariant manifolds associated to nonzero
Lyapunov exponents in non-uniformly hyperbolic systems: this extends to a larger class of
dynamical models the study of Lyapunov stability for a measure--preserving dynamics.
The main achievements by Pesin are contained in \cite{pesin2}. In that manuscript
he proved that the Kolmogorov--Sinai entropy $K(L)$ of  any classical dynamical system 
$(\mtM,\mu,L)$ can be expressed in terms of the LE, through the relation
\be
\label{KS}
K(L) = A \int_{\mtM} \sum_{i}^+ \lambda_i(x) \mathrm{d}\mu \, ,
\ee  
where $A$ is a suitable constant and the sum $ \sum_{i}^+$ is restricted to the positive
$\lambda_i$.
One of the important consequences of this result for practical applications is that
the entropy of a dynamical system (one could say, its {\it degree of unpredictability})
has a density with respect to the measure $\mu$ and this density is related to the
sum of the positive LE's. For an ergodic system, since the LE are constant almost-everywhere
the above expression becomes
\be
\label{KS2}
K(L) = A \sum_{i}^+ \lambda_i
\ee  
In the same paper \cite{pesin2} Pesin discussed how the mechanism of Oseledets' splitting
(i.e. the existence of a filtration of the tanget--space evolution into invariant submanifolds)  is
related to the hyperbolicity of the dynamical system
$(\mtM,\mu,L)$.\\ 

\section{The geometrical structure of tangent space}
\label{sec3}
The existence for invertible dynamics of a coordinate--independent
local decomposition of the tangent--space evolution into covariant subspaces (again,
the mechanism of Oseledet's splitting) was discussed by Ruelle in a seminal paper of
1979 \cite{ruelle}. Rather than surveying the rigorous mathematical treatment, we have
decided to describe here, in a language accessible also to non mathematicians, the main
consequences of Ruelle's contribution. For this purpose we shall adopt 
notations more familiar to an audience of readers
interested to applications. In practice, rather than referring to  an  abstract dynamical system, 
$(\mtM,\mu,L(x))$, we shall consider explicitly models of physical interest, where  $\mtM$ is identified with
$\R^N$ and the evolution operator $L(x)$ is
specified by a set of ordinary differential equations
\be
\label{dync}
\dot{\bx}_t = \bof(\bx_t)
\ee
where the point $x\in \mtM$ is now represented by the vector $\bx_t 
= \{x^1_t,x^2_t, \cdots , x^N_t \}\in \R^N$ and the continuous 
dependence on time is made explicit by the subscript $t \in \R$, so that 
the application  ${\bof}:\R^N \to \R^N$ is a continuous--time dynamical rule. In this section we limit
our analysis to invertible dynamical rules, i.e. we assume that for any given position
in the phase-space there exists only one backward trajectory.
Since a continuous-time dynamics can be reduced to a discrete-time one with the
help of a Poincar\'e section \cite{GH}, we make reference only to the map
formalism, where the dynamics reads
\be
\label{dynd}
\hspace {-1.cm}\bx_m = \bof^{(m)}(\bx_0) \quad \quad , \quad \bx_{m+n} = \bof^{(m)}(\bx_n) = 
\bof^{(n)}(\bx_m)  = \bof^{(m)}  \bof^{(n)}(\bx_0)
\ee
where $\bx_m= \{x^1_m,x^2_m, \cdots , x^N_m\}\in \R^N$ and 
the time variable is an integer index $m \in
\mathbb{Z}$. Again, the map $\bof^{(m)}$ is assumed to be invertible, i.e.
the inverse map $\bof^{(-m)}(\bx_m)$ exists and has only one solution.

It is worth pointing out that the assumption of the existence of an invariant measure $\mu$ is a crucial point that
demands much care to be properly formulated\footnote{In the
  case of symplectic evolution, ergodicity is
usually assumed, in spite of examples where it is known that the ergodic
hypothesis does not hold (e.g., the Fermi--Pasta--Ulam model). Moreover, even 
uniform hyperbolicity or simply hyperbolicity are not guaranteed on rigorous grounds
for many models of interest for applications. This notwithstanding,  many of the 
results and tools derived from rigorous ergodic theory reveal effective and very useful, 
when applied to dynamical models that, presumably, do not fullfill all the hypotheses from
which they were derived, as captured by the celebrated chaotic
hypothesis\cite{chaohyp}. 
Anyway, one should always keep in mind that a straightforward
application of these methods may yield ``pathological'' outcomes that are just a 
consequence of singular features emerging from the highly complicated nature of 
$\mu$, provided it does exist.}.
In this discrete framework it is convenient to consider the Jacobian 
matrix at $\bx_n$
\be
{\bf J}(\bx_n) = \frac{\partial \bof(\bx_n)}{\partial \bx_n}  \quad ,
\ee
whose elements are assumed to be non-divergent at any $\bx_n$ 
and defined as follows
\be
[{\bf J}(\bx_n)]_{i,j} = \frac{\partial (x^{(i)}_{n+1})}{\partial x^{(j)}_n}
\ee
with $x^{(i)}_n=[\bx_n]_i$.
The evolution operator in tangent space $D L^t(x)$ introduced in Eq.~(\ref{symts}) 
can be represented by the matrix 
\be
{\bf M}_{ k,n} = \prod_{i=n}^{k+n-1}{\bf J}({\bx_{i}})
\ee
where  ${\bf M}_{ k,n} : \R^N \to \R^N$ and the lapse of time $t$ has become the integer $k$.
Note also that with a slight abuse of notation we write ${\bf M}_{ k,n}$ instead of ${\bf M}_{k}({\bf x_n})$.\\ 
It can be easily shown  that ${\bf M}_{k,n}$ satisfies cocycle properties 
\be
\hspace {-1.cm}{\bf M}_{k+l,n} = {\bf M}_{k,n+l}{\bf M}_{l,n} \quad , \quad {\bf M}_{0,n} = \mathbb{I}
\quad \Rightarrow \quad {\bf M}_{-k,n} = ({\bf M}_{k,n-k})^{-1}\,.
\ee
The first relation defines the multiplicative property of Jacobian matrices, while the last one
defines the inverse, i.e. time--reversed, evolution matrix in the tangent space.

\subsection{Oseledets' matrix and Oseledets' splitting}
\begin{figure}
\centering
\includegraphics[draft=false,clip=true,width=0.45\textwidth]{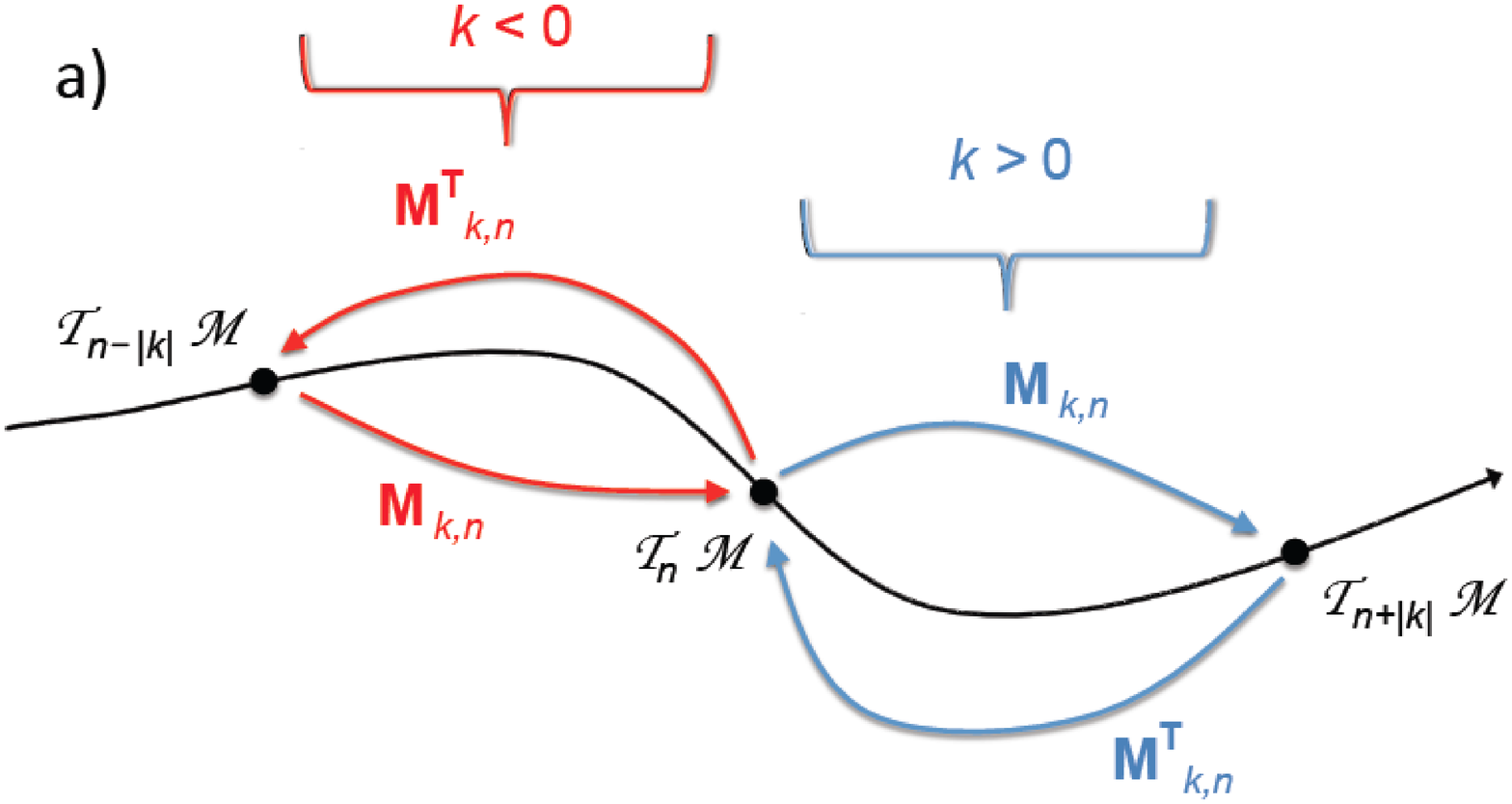}
\includegraphics[draft=false,clip=true,width=0.45\textwidth]{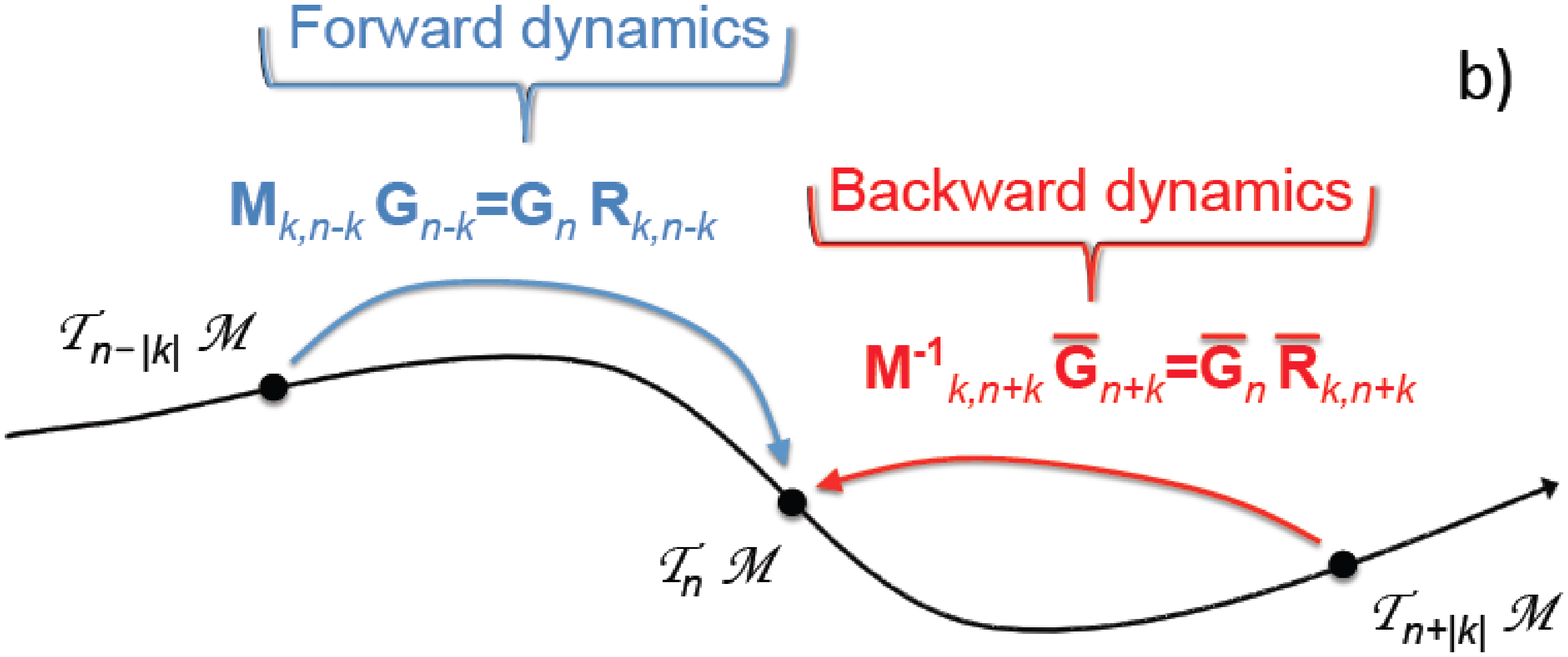}
\caption{(color online) (a) Schematic representation of the action of the
  symmetric matrices ${\bf M}^T_{k,n} {\bf M}_{k,n} $ on tangent space $\mtT_n \mtM$
  for finite $k$. (b) Forward ($+$) and backward ($-$) Gram-Schmidt evolution
  (\ref{split1}) in tangent space for finite $k$. The oriented black
  line represent the phase space dynamics (\ref{dync}). Blue arrows represent the forward
  dynamics, red ones the time reversed, backward one. } 
\label{fig1} 
\end{figure}

It is worth reformulating Oseledets' theorem by
taking advantage of the invertible dynamics and introducing the concepts
of forward and backward Oseledets' matrix, denoted by the signs $"+"$ and $"-"$,
respectively.  Under the same hypotheses as for Oseledets' theorem, one can prove
that the following limits exist
\be
\label{osma}
{\bf \Xi}^{\pm}_{n}=\lim_{k \pm \infty} \frac{1}{2 \,k} \ln
\left[ ({\bf M}_{k,n})^T {\bf M}_{k,n} \right]
\ee
where $T$ indicates the transpose matrix\footnote{We assume that
  the scalar product is the one associated with the chosen basis. In
  order to consider a more generic scalar product, one should use
  the adjoint matrix $({\bf M}_{k,n})^*$ instead of $({\bf M}_{k,n})^T$.}.
Note that for $k>0$ ($k<0$) the matrix $ ({\bf M}_{k,n})^T {\bf
  M}_{k,n}$ evolves forward (backward) a generic tangent space vector
from time $n$ to time $n+k$,
and then backward (forward) in time up to time $n$. Therefore, the
forward Oseledets' matrix ${\bf \Xi}^+_n$ probes the future dynamics of ${\bf x}_n$,
while the backward one ${\bf \Xi}^-_n$ probes the past (see Fig. \ref{fig1}).\\
The forward and the backward Oseledets' matrices share the same $m\le
N$ distinct eigenvalues 
$\lambda_1 > \lambda_2 > \ldots > \lambda_m$, each with multiplicity
$g_1$ ($\sum_{i=1}^N g_i=N$) which coincide with the LEs of the dynamical system (\ref{dync}).
The orthogonal eigenvectors $({\bf d}^{(i)}_n)_{\pm}$ (with $i=1,\ldots,N$) of 
${\bf \Xi}^{+}_n$ and ${\bf \Xi}^{-}_n$, on the other hand, do differ and are not invariant under time reversal.
As shown in \cite{ruelle}, one can construct 
Oseledets' splitting ${\bf \Omega}^{(i)}_n$ ($i=1,\ldots,m$) by simply intersecting the so--called Oseledets' subspaces 
\be
\label{ome}
{\bf \Omega}^{(i)}_n = ({\bf \Gamma}^{(i)}_n)^{+}
\cap ({\bf \Gamma}^{(i)}_n)^{-}
\ee
where 
\be
\label{ose1}
({\bf \Gamma}^{(i)}_n)^{+} = ({\bf U}^{(i)}_n)^{+}\oplus
\ldots \oplus ({\bf U}^{(m)}_n)^{+}
\ee
and
\be
\label{ose2}
({\bf \Gamma}^{(i)}_n)^{-} = ({\bf U}^{(1)}_n)^{-}\oplus
\ldots \oplus ({\bf U}^{(i)}_n)^{-}
\ee
In both formulae $({\bf U}^{(i)}_n)^{\pm} $ with $i=1,\ldots,m$ denote the eigenspaces of 
Oseledets' matrices defined in  (\ref{osma}), i.e. they are the
(orthogonal) subspaces spanned by the eigenvectors of the forward and
backward Oseledets' matrices.
Note that, as  a consequence of Oseledets' theorem,  the subspaces $({\bf \Gamma}^{(i)}_n)^{\pm}$
are a filtration of the dynamics (\ref{dync}) or of its time-reversal, that is
\be
\label{splitting1}
\lim_{k \to \pm \infty} \frac{1}{|k|} \ln \frac{\|{\bf M}_{k,n} {\bf
    u}\|}{\|{\bf u}\|}=\pm \lambda_i
\;\;\;\mathrm{for} \;\;{\bf u} \in ({\bf \Gamma}^{(i)}_n)^{\pm} \setminus ({\bf \Gamma}^{(i \pm 1)}_n)^{\pm}
\ee  
with the nested subspace structure $\R^N=({\bf \Gamma}^{(1)}_n)^{+}\supset ({\bf \Gamma}^{(2)}_n)^{+} \supset \ldots 
\supset {\bf (\Gamma}^{(m)}_n)^{+}\supset {\bf (\Gamma}^{(m+1)}_n)^{+}
\equiv \emptyset$
and $\R^N=({\bf \Gamma}^{(m)}_n)^{-}\supset ({\bf \Gamma}^{(m-1)}_n)^{-} \supset \ldots 
\supset {\bf (\Gamma}^{(1)}_n)^{-} \supset {\bf (\Gamma}^{(0)}_n)^{-}
\equiv \emptyset$. 
%
%
Oseledets' splitting $\{{\bf \Omega}^{(i)}_n\}_{i=1,\ldots,m}$ is
generically non-orthogonal, and determines a {\it measurable
decomposition} of the tangent
space which is independent of the chosen norm, covariant with the
dynamics and (obviously)
invariant under time reversal, that is
${\bf M}_{k,n} {\bf \Omega}^{(i)}_n={\bf \Omega}^{(i)}_{k+n}$. 
The Oseledets' decomposition into such {\it covariant} subspaces exists for any map (\ref{dynd}),
which is continuous and measurable together with its inverse and whose Jacobian matrix exists
and is finite in each element. If the Jacobian matrix is
non-degenerate ($m=N)$,
it can be easily shown that also the spectrum of LEs is non
degenerate and all the subspaces ${\bf \Omega}^{(i)}_n$ have dimension
one. Conversely, if the Jacobian matrix is degenerate, 
each covariant subspace has a dimension equal to the degeneracy $g_i$ of the corresponding LE.\\
Finally, as a consequence of Eqs.~(\ref{ome},\ref{splitting1}), it easy to show
that a generic tangent space vector, ${\bf u}^{(i)}_n \in {\bf \Omega}^{(i)}_n$
for large values of $k$ grows as  
\be
\label{clv-lambda}
\| {\bf M}_{k,n} {\bf u}^{(i)}_n\| \sim \|{\bf u}^{(i)}_n\| \exp^{\lambda_i k}\,.
\ee
that is, its exponential asymptotic growth rate is the LE $\lambda_i$.

\subsection{Gram--Schmidt vectors and Lyapunov characteristic exponents}
The existence of a well defined limit in Eq.~(\ref{osma}) does
not allow one to compute neither the spectrum of the Lyapunov
characteristic exponents
nor the Oseledets' subspaces and splitting by directly diagonalizing the
symmetric matrices ${\bf \Xi}^{\pm}_n$, since they develop
exponentially diverging terms leading to numerical overflows. A way
out, as noted in Refs. \cite{Shimada, Benettin}, is offered by
Eq.~(\ref{lambdap}). The idea is to follow the tangent space evolution of 
$p$-volumes of increasing dimension $p$, and to recover the LEs
$\lambda_i$ from Eq.~(\ref{lambdap}). While $p$-volumes may be defined
by an arbitrary basis of $p$ independent vectors 
$(\tilde{\bf g}^{(1)}_n, \tilde{\bf g}^{(2)}_n, \ldots, \tilde{\bf
  g}^{(p)}_n)$, in order to evolve
volumes of dimension $p>1$, one has to take into account the
fact that
almost every initial vector will converge exponentially towards the
largest expanding one,
\be
\label{tildebasis}
\|{\bf M}_{k,n} \tilde{\bf g}^{(i)}_n - {\bf M}_{k,n} \tilde{\bf
  g}^{(j)}_n\| \to 0\;\;\;\mathrm{for}\;\;k\gg1\;\;\mathrm{and}\;\forall\, i,j\,,
\ee 
thus quickly making any two vectors $i$ and $j$ numerically undistinguishable.
To solve this problem it is necessary to periodically orthogonalize
the tangent space basis during time evolution. If one considers an
orthogonal matrix whose vector columns form an orthonormal basis in tangent
space, ${\bf G}_n=\left({\bf g}^{(1)}_n|{\bf g}^{(2)}_n|\ldots|{\bf g}^{(N)}_n\right)$,
this can be achieved by means of QR decomposition every $k$ timesteps,
\be
\label{split1}
{\bf M}_{ k,n} {\bf G}_n = {\bf G}_{k+n} {\bf R}_{k,n} 
\ee
where ${\bf G}_{k+n}$ is made of orthonormal vectors at $\bx_{k+n} = \bof^{(k)} (\bx_n)
= \bof^{(k+n)} (\bx_0)$ and the upper diagonal matrix ${\bf R}_{k,n}$
contains the information obtained by the orthonormalization procedure
of the {\it non-orthogonal} matrix $\tilde{\bf G}_{n+k}={\bf M}_{ k,n} {\bf
  G}_n$. In particular, its diagonal elements $[{\bf
  R}_{k,n}]_{i,i}=\gamma^{(i)}_{k,n}$ are the local growth
rates over a time $k$ of the orthogonal vectors ${\bf G}_n$, that is
the growth rates of the axes of inertia of $p$-volumes. 
In numerical applications, $k$ should not be so large as to produce overflows or problems of
numerical accuracy in the QR decomposition of $\tilde{\bf G}_{n+k}$.
In systems
with an invariant ergodic measure $\mu$ on $\mtM$, where 
averages over $\mu$ can be replaced by time averages, the (ordered) Lyapunov
spectrum can be thus computed by taking the time-average of the logarithms
of these diagonal terms,
\be
\label{lyapB}
\lambda_i=\lim_{T \to \infty} \frac{1}{T} \sum_{h=0}^{T-1}
\ln \gamma^{(i)}_{k,n + h k}\,.
\ee
The off-diagonal nonzero elements $[{\bf R}_{k,n}]_{j,i}$, on the
other hand, are obtained
by projecting each vector column $\tilde{\bf g}^{(i)}_{n+k}$ onto 
the subspace spanned by $\{\tilde{\bf g}^{(j)}_{n+k}\}$ with $j<i$.\\
The orthogonal column vectors of ${\bf G}_{m}$ evolving under Eq.~(\ref{split1})
are often referred to as {\it forward} Gram-Schmidt (GS) vectors (from the well-known
orthonormalization algorithm). In general, they depend on the initial condition $\bx_n$
from which the above procedure has been initialized, but for $m-n$ sufficiently large they
converge to a well defined direction which only depends on $m$. \\
This is the main result obtained by Ershov and Potapov \cite{EP}: 
under the general hypothesis that the considered dynamical system has an
invariant ergodic measure, one can show that the orthonormal basis
$\{{\bf g}^{(i)}_{n}\}_{i=1,\ldots,N}$ evolved by Eq.~(\ref{split1})
converges exponentially to the eigenvectors $\{({\bf d}^{(i)}_n)_{-}\}_{i=1,\ldots,N}$ of the
{\it backward} Oseledets' matrix,
\be 
\|{\bf g}^{(i)}_{n+k} - ({\bf d}^{(i)}_{n+k})_-\| \to 0\;\;\;
\mathrm{for} \;\;k \to \infty\,.
\ee
From now on we will always refer to GS vectors as to the ``asymptotic" ones\footnote{
Ershov and Potapov call the large time limit of the GS vectors a 
{\it stationary basis}, a somewhat confusing term since they do vary
in time. We therefore prefer to use the term ``asymptotic''.}. \\

An analogous result holds for
the {\it backward} Gram-Schmidt vectors of the time reversed
dynamics, which converge 
to the same index eigenvectors of the {\it forward} Oseledets' matrix. Obviously,
the (reversely ordered) LEs $\bar{\lambda}_i$ of the time reversed dynamics coincide
with the forward ones apart for a sign change
\be
\lambda_i=-\bar{\lambda}_{i}
\ee
and can be numerically computed by the analogous of
Eq.~(\ref{lyapB}). Note that by this convention, the fastest growing
direction of the time-reversed dynamics is associated to the $N$-th
backward GS vector.\\
It is also important to stress that 
{\it the forward and backward Gram-Schmidt vectors are not invariant under time-reversal}.
Furthermore, they are norm-dependent objects -- as it follows from the
norm-dependent nature of GS orthogonalization -- with the exception of the
first forward and first backward GS vectors (i.e. the ones associated to the
largest expansion rate in both time directions).\\
The fact that the forward GS vectors converge to the
eigenvectors of the backward Oseledets' matrix could seem
odd at first sight\footnote{In order to avoid this ``inversion'' between 
 Oseledets eigenvectors and asymptotic GS vectors, some author indeed prefer to name
  ``backward'' the GS vectors obtained by moving forward in time, and
  viceversa. However, since we prefer to emphasize the operational way
  used to compute them, i.e. by following the forward dynamics, we
  will stick with our nomenclature.} but it is easy to realize that both basis in $\mtT_n \mtM$
are only determined by information pertaining to the past dynamics of ${\bf x}_n$, as
shown in Fig. \ref{fig1}. Once again, the same heuristic reasoning
applies to backward GS vectors and the eigenvectors of the forward Oseledets' matrix.

\section{Covariant Lyapunov vectors}
\label{Sec4}
The possibility to characterize and practically compute a
norm-independent and 
time-invariant set of local tangent space vectors associated to the
LEs allows one to directly probe the expanding and contracting directions
of a given dynamical system. This development has a number of noteworthy applications
beyond the area of ergodic theory, ranging from ensemble forecasting to
statistical mechanics. It is thus natural to consider the
{\it covariant Lyapunov vectors} ${\bf v}^{(i)}_n$, which are
the unitary vectors spanning the Oseledet's splitting ${\bf \Omega}^{(i)}_n$. \\
The knowledge of CLVs allows to identify at each point in phase space a vector field that has
a natural geometrical interpretation: it is the collection of the
tangent space varieties associated to the stability properties of the dynamical system ruled 
by the LE's. Accordingly, CLVs are 
objects of primary importance for the study of a dynamical system, since they provide
all the information about the local geometrical structure of the
tangent space. We now discuss them in detail.\\
If  $\mathrm{dim} ({\bf \Omega}^{(i)}_n) = 1$, that is, if $\lambda_i$
is non-degenerate, a single CLV ${\bf v}^{(i)}_n$ associated to the LE
$\lambda_i$ is uniquely defined (up to a phase factor), while for degenerate LEs any
nonsingular base formed by $g_i=\mathrm{dim} ({\bf \Omega}^{(i)}_n)$
covariant vectors
${\bf v}^{(i)}_n \in {\bf \Omega}^{(i)}_n$ may be arbitrarily
considered. In the following, unless otherwise stated, we will unfold eventual degeneracies of
the Lyapunov Spectrum, and consider $N$ LEs $\lambda_1 \ge \lambda_2
\ge \ldots \ge \lambda_N$ and their corresponding CLVs $\{ {\bf v}^{(i)}_n\}_{i=1,2,\ldots,N}$.

In principle, CLVs can be obtained by: i) a ``dynamical" approach, 
which consists in first determining the Lyapunov basis via the GS dynamics 
(\ref{split1}) forward in time and and then following the time-reversed evolution
in the corresponding subspaces \cite{Ginelli2007}) ; ii) a ``static" approach which consists in computing 
the corresponding Oseledets's subspaces via (\ref{ose1}),(\ref{ose2}) and subsequently 
intersecting them according to Eq.~(\ref{ome}) \cite{Legras}. 
The intersection of large matrices may be susceptible to problems of numerical 
accuracy in high-dimensional spaces. Although some variants of the ``static'' 
approach, have been recently proposed \cite{WS,KP}, which improve the performance
of such a method, we believe that the former ``dynamical" approach is more
efficient and we thereby provide a detailed description in following subsection
and discuss its stability in Section~\ref{sec5}. Anyway, for the sake of completeness,
in Section~\ref{sec5}, we briefly illustrate also the static algorithms. 

\subsection{Dynamical algorithm for computing covariant Lyapunov
  vectors - Formal aspects}
The dynamical algorithm sketched in \cite{Ginelli2007} makes use of the following simple but powerful
observation. Suppose we have computed the long-time limit of the forward GS vectors (i.e. the
asymptotic GS basis) along a certain trajectory. 
Call these GS vectors ${\bf g}^{(i)}_n$. Then, the $j$-th CLV is contained  
in the subspace formed by the first
$j$ asymptotic GS vectors. Moreover, a generic vector ${\bf u}
\in \R^N$ evolved backward from time $n$ through the time-reversed dynamics
within the corresponding Oseledets' subspace $({\bf \Gamma}^{(j)}_m)_-$ will converge asymptotically 
to the true CLVs ${\bf v}^{(j)}_m$.
The very fact that CLVs are computed via a stable dynamical rule and not by intersections
of large subspaces ensures the numerical stability of this algorithm (see also the next
section).

Our first observation follows trivially from the definitions
(\ref{ome}-\ref{ose2}). 
It can be expressed as 
\be
\label{clv-exp}
{\bf v}^{(i)}_n = \sum_{j=1}^i c_n^{(j,i)} {\bf g}^{(j)}_n
\ee
where the $c_n^{(j,i)}=\langle{\bf g}^{(j)}_n | {\bf v}^{(i)}_n\rangle$ with $j\le i$ are the CLVs expansion coefficients at time $n$.
Note that CLVs are defined up to an irrelevant sign (i.e. only their
orientation is defined)
and that the first CLV coincides with the first vector of the
GS basis by construction.

To demonstrate the second point, we first express
the CLVs dynamics in a convenient form. Define 
the matrix whose vector columns are the CLVs, 
${\bf V}_n=\left({\bf v}^{(1)}_n|{\bf v}^{(2)}_n|\ldots|{\bf
    v}^{(N)}_n\right)$.
Covariant vectors evolve according to
\be
{\bf M}_{k,n} {\bf V}_n = {\bf V}_{n+k} {\bf D}_{k,n}
\label{clv1} 
\ee
where the diagonal matrix ${\bf D}_{k,n}$ is composed of the
local growth factors $\gamma^{(i)}_{k,n} = \| {\bf M}_{k,n} {\bf v}^{(i)}_n
\|$, that is $[{\bf D}_{k,n}]_{i,j} = \delta_{i,j} \gamma^{(i)}_{k,n}$.
For finite $k$, the logarithms of these growth factors are called
{\it finite time} Lyapunov exponents (FTLE)
\be
\lambda_i^{(k,n)}=\ln \gamma^{(i)}_{k,n}
\label{FT}
\ee
and their time average obviously coincide with the LEs.

For the sake of simplicity, we assume that the spectrum is not degenerate, so
that all individual CLVs may be resolved, and 
we express Eq.~(\ref{clv-exp}) in matrix form. We write ${\bf
  V}_n={\bf G}_n {\bf C}_n$, where the upper triangular matrix 
${\bf C}_n$ contains the  CLVs expansion coefficients, 
$[{\bf C}_n]_{j,i}=c_n^{(j,i)}$ for $j \le i$. Note that since the
CLVs have unit norm by definition, the expansion
coefficients have to be normalized column by column, 
\be
\label{norm-clv}
\sum_{j=1}^i (c_n^{(j,i)})^2=1 \;\;\forall\;i\,.
\ee
Using Eq.~(\ref{split1}) we may rewrite Eq.~(\ref{clv1}) as
\be
{\bf G}_{n+k} {\bf C}_{n+k} {\bf D}_{k,n} = {\bf M}_{k,n} {\bf G}_n {\bf C}_n = {\bf G}_{k+n} {\bf R}_{k,n} {\bf C}_n
\label{clv2} 
\ee
 which implies
\be
{\bf C}_n = {\bf R}_{k,n}^{-1}  {\bf C}_{n+k} {\bf D}_{k,n}\,.
\label{clv-dyn}
\ee
This is our fundamental equation for backward evolution, showing that
the CLVs -- once expressed in the basis 
of the GS vectors -- can be easily evolved backward by multiplying the inverted upper triangular
matrices ${\bf R}_{k,n}^{-1}$, where ${\bf R}_{k,n}$ is a ``by-product'' of the forward GS
dynamics (\ref{split1}).

Now consider a generic nonsingular upper triangular matrix  $\widetilde{\bf C}_{n+k}$ which,
in the GS basis, defines a nonsingular sets of vectors 
$\{ \tilde{\bf u}^{(i)}_{n+k}\}_{i=1,\ldots,N}$ projected in the
Oseledets' subspaces, such that 
$\tilde{\bf u}^{(i)}_{n+k} \in ({\bf \Gamma}^{(i)}_{n+k})_-$.
We compare its backward evolution with that of true CLVs, ${\bf u}^{(i)}_{n+k}$
(which in the GS vectors basis we write in the matrix form ${\bf C}_{n+k}$).

Evolving both sets of vectors backward according to Eq.~(\ref{clv-dyn}), we have
\begin{eqnarray}
\widetilde{\bf C}_n &=& {\bf R}_{k,n}^{-1}  \widetilde{\bf C}_{n+k} 
\widetilde{\bf D}_{k,n}\\
{\bf C}_n &=& {\bf R}_{k,n}^{-1}  {\bf C}_{n+k} {\bf D}_{k,n}
\end{eqnarray}
which implies
\be
\widetilde{\bf C}_{n+k} \widetilde{\bf D}_{k,n} \widetilde{\bf C}_n^{-1} =
{\bf C}_{n+k} {\bf D}_{k,n}  {\bf C}_n^{-1} 
\ee
and thus
\be
\widetilde{\bf C}_{n}^{-1} {\bf C}_{n} = 
\widetilde{\bf D}_{k,n}^{-1} \widetilde{\bf C}_{n+k}^{-1}
{\bf C}_{n+k} {\bf D}_{k,n}\,.
\ee
It is easy to verify that 
\be
[\widetilde{\bf C}_{n}^{-1} {\bf C}_{n}]_{i,j}
=\frac{\gamma_{k,n}^{(j)}}{\tilde{\gamma}_{k,n}^{(i)}} [\widetilde{\bf C}_{n+k}^{-1}
{\bf C}_{n+k}]_{i,j}
\label{components}
\ee
where both sides of Eq.~(\ref{components}) are upper triangular
matrices. Thus, the r.h.s. matrix-elements are zero for $i>j$. On the other
hand, for $i<j$ we
have
\be
\frac{\gamma_{k,n}^{(j)}}{\tilde{\gamma}_{k,n}^{(i)}} \to
0\;\;\;\mathrm{for}\;\; k\to\infty \;\;\mathrm{and}\;\;i<j
\label{clv-conv}
\ee
due to the nested structure of the Oseledets' subspaces. For
$k\gg1$ one has (up to an irrelevant sign)
\be
[\widetilde{\bf C}_{n}^{-1} {\bf C}_{n}]_{i,j}
\approx \delta_{i,j} [\widetilde{\bf C}_{n+k}^{-1}
{\bf C}_{n+k}]_{j,j} \equiv \delta_{i,j} q_j
\label{components2}
\ee
where $q_j$ are arbitrary factors which solely depend on the angles between the initial
vectors at time $n+k$. We finally have, for $k\gg1$,
\be
[\widetilde{\bf C}_{n}]_{i,j}
\approx [{\bf C}_{n}]_{i,j}\,q_j^{-1}\,
\ee
that is, under the backward evolution (\ref{clv1}) 
almost any nonsingular vector set (once it belongs to the Oseledets' subspace)
converges backward in time towards the true CLV, apart from a trivial scale factor. 
This proves our second point.

As a final practical remark, we want to point out that numerical estimates of LE and CLV are always performed
over long but obviously finite time lapses, whereas all theorems hold in the infinite time limit. 
In the next Section, we show that the convergence rate is typically exponential, though the
rate may be rather small for large system sizes. Accordingly, we do not expect any practical problem.

\subsection{Dynamical algorithm for computing covariant Lyapunov
  vectors - Numerical implementation}
\label{Sec4B}

We now discuss the numerical aspects of the dynamical algorithm. The first part closely follows 
the standard algorithm introduced by Benettin {\it et al.} to compute LEs \cite{Benettin}, while in the
second part, backward evolution is implemented via Eq.~(\ref{clv1}).
\begin{enumerate}
\item
{\it Forward transient} --
Given a generic initial condition ${\bf x_0}$ in phase space and a generic set of $m \le N$ orthogonal 
tangent-space vectors $\{ {\bf g}^{(i)}_0\}_{i=1,\ldots,m}$, we first evolve the
phase and the tangent space dynamics via Eq.~(\ref{dync}) and
Eq.~(\ref{split1}), respectively. This transient should last a
number $n$ of timesteps sufficient for the phase-space
trajectory to converge to the ergodic attractor and for the orthogonal
vectors to converge to the asymptotic GS vectors.
\item
{\it Forward dynamics} --
Once a GS basis is reached, we proceed further
evolving the reference trajectory ${\bf x}_n$ and GS vectors by $k$ timesteps at once, recording in
memory both the local GS vectors ${\bf G}_{n+hk}$ and the
upper triangular matrices ${\bf R}_{k,n+(h-1)k}$, for $h=1,2,\ldots,t+t_0$.

\item
{\it Backward transient} --
A generic non-singular upper triangular matrix $\widetilde{\bf
  C}_{n+(t+t_0)k}$ is generated and evolved backward via
Eq.~(\ref{clv1}) for $t_0$ steps (each moving backward by $k$
timesteps). The backward transient length $k t_0$ should
be sufficient to converge the tangent space initial conditions close
enough to the true CLVs expansion coefficients in ${\bf x}_{n+t k}$.
Columns of $\widetilde{\bf
  C}_{n+(t+t_0)k}$ are kept normalized via the condition (\ref{norm-clv}).

\item
{\it Backward dynamics} --
Finally, the CLVs expansion coefficients may be further evolved
backward along the trajectory 
(always being kept normalized) to sample the geometrical structure of the ergodic attractor at 
points ${\bf x}_{n+h k}$, with $h=0,1,\ldots,t$.
CLVs expressed in the phase space coordinates reference can be
obtained by the GS vectors ${\bf G}_{n+hk}$ via Eq.~(\ref{clv-exp}).

\end{enumerate}
Several comments are in order:
\begin{itemize}

\item
{\it Sampling frequency} --
Note that $k$ dictates the sampling frequency of CLVs over the phase space, reference trajectory 
${\bf x}_{n+h k}$.  It can be adjusted as needed, but it should be small enough to avoid numerical 
overflows and too large numerical errors in the ${\bf R}_{k,m}$ and ${\bf C}_{m}$ matrices due to 
exponential growth and contractions. The choice of $k$ thus obviously depends on both the stability 
properties of the chosen dynamical system and on the required sampling frequency. In principle, two 
different values can be chosen, one ($k_{tr}$) for the forward and backward transients, and a second 
$k_{dyn}<k_{tr}$ for the forward and backward dynamics.

\item
{\it Scalability} --
The algorithm is fully scalable, and if one just needs to compute the
first $m \le N$ CLVs, that is the $m$ most expanding vectors, no other
vectors with $i>m$ are needed.
Indeed, in order to compute the first $m$ CLVs one only needs to consider the $N\times m$ reduced GS 
matrices ${\bf G}_n$ and the corresponding $m\times m$ reduced upper triangular matrices ${\bf R}_n$. 
If the dynamics is explicitly invertible, the same reasoning could be applied to the computation of 
the last $m$ CLVs, i.e. the $m$ most contracting vectors, by simply writing the dynamical algorithm 
for the time-reversed dynamics (provided that the right trajectory has been generated forward in time).

\item
{\it Degenerate Lyapunov spectra} --
If the Lyapunov spectrum is degenerate, and some covariant subspaces ${\Omega}^{(i)}_m$ have a dimension $g_i$ larger than one, 
the individual vectors spanning such a degenerate subspace have no physical meaning.
In this case, the above algorithm will simply return some arbitrary basis of ${\Omega}^{(i)}_m$ which may depend on the initial choice of 
$\widetilde{\bf C}_{n+(t+t_0)k}$. For a more detailed discussion of the structure of the degenerate CLVs associated to zero LE 
in a quasi-one dimensional Hamiltonian systems of hard disks see for instance Ref. \cite{Morriss11}.

\item
{\it Memory issues} --
All existing algorithms need to store a large amount of information. 
Our dynamical algorithm requires to store the ${\bf R}_{k,m}$ and ${\bf G}_{m}$ matrices 
at periodic sampling points during the forward dynamics, in order to be reused during 
the backward evolution. The matrices ${\bf R}_{k,m}$ needed to run the backward dynamics involve
$m(m+1)/2$ floating point numbers at each of the $t+t_0$ sampling points, while the ${\bf G}_{m}$ matrices 
involve $mN$ floating-point numbers for $t$ sampling points. In terms of floating-point numbers, the
total memory requirement is thus
\be
M_{tot}=t\,m\,\left[N+\frac{m+1}{2}\right]+t_0\frac{m(m+1)}{2}\,.
\ee 
This burden can be sensibly reduced if one is not interested in the spatial structure of CLVs but just 
in their relative angles, since the ${\bf G}_{m}$ matrices are not needed (the relative angles can be 
computed directly with reference to the GS vectors basis). In this case, the memory requirement is just
\be
M'_{tot}=(t+t_0)\,\frac{m(m+1)}{2}\,.
\ee
These estimates could nevertheless turn into a large amount of memory for long sampling times, large dynamical 
systems and/or when many CLVs are required, quickly exhausting the available fast-access memory. 
Since accessing disk storage every $k$ timesteps could be rather time-consuming (and disk memory in itself 
is a limited resource), we suggest an alternative strategy. Once determined the maximum amount of 
data $M_b$ that can be stored in the fast-access memory, divide the total forward integration time $k(t+t_0)$
into $n_b$ blocks of length $h_b\,k$ such that $M_{tot}/n_b \leq M_b$. 
After running the usual forward transient in order to ensure a proper convergence of the GS vectors 
at time $0$, perform a first forward-dynamics run, without storing the ${\bf R}_{k,m}$ and ${\bf G}_{m}$ 
matrices every $k$ timesteps as usual, but just saving the current phase space configuration
and the GS vectors every $h_b\,k$ timesteps at times $n h_b k$, with $n=0,1,\ldots,n_b-1$, 
that is, at the {\it beginning} of each block (this means one does not have to run forward in time the 
last block). These data can be typically stored on disk, as it will not be accessed too frequently.
Once the first forward run is concluded, perform a series of forward and backward iterations block-by-block, 
{\it starting from the last block}. Thus, recover the phase space trajectory and GS vectors at time 
$(n_b-1) h_b k$, evolve them forward up to time $n_b h_bk$, now storing in the fast access memory the 
${\bf R}_{k,m}$ and ${\bf G}_{m}$ matrices every $k$ timesteps, and subsequently evolve backward a randomly 
generated nonsingular initial condition $\widetilde{\bf C}_{n_b h_b k}$.
Next, pass to the $(n_b-1)$-th block, accessing the data stored at time $(n_b-2) h_b k$, performing the 
forward evolution up to time $(n_b-1) h_b k$. When coming back, however, restart from the coefficient 
computed in the previous block, $\widetilde{\bf C}_{(n_b-1) h_b k}$, so that 
the convergence to the true CLVs is not lost. This procedure is finally repeated block by block
up to the first one, greatly reducing the amount of memory, at the only expense of performing twice the
same forward and backward dynamics, thus doubling the algorithm computational time.

\item
{\it Computational time} --
We give order-of-magnitude estimates for the above dynamical algorithm in terms of elementary 
arithmetic operations (basically multiplication/division floating point operations).
Suppose we have a dynamical system with $N$ degrees of freedom and we are interested in computing 
the first $m$ CLVs.

Forward dynamical evolution requires to run both phase and tangent space dynamics for $k$ timesteps 
and then to perform a single QR decomposition (or Gram-Schmidt orthonormalization). Special dynamical 
systems such as the ones with strictly local or globally coupled interactions (in the following: 
easy dynamics)  require only ${\cal O}(N)$ operations for a single step of phase space dynamics and 
${\cal O}(mN)$ for the tangent space evolution. Unfortunately, generic dynamical systems with long-range 
interactions (in the following: hard dynamics) typically require ${\cal O}(N^2)$ and ${\cal O}(mN^2)$ 
operations for phase-space and tangent-space 
dynamics\footnote{We assume the usage of standard matrix multiplication algorithms.} 
respectively (the latter with a prefactor $\leq 1$). 
This leads to ${\cal O}(m\,k\,N)$ (for easy dynamics) or ${\cal O}(m\,k\,N^2)$
(for hard dynamics) operations to be performed between consecutive QR decompositions, 
which is an ${\cal O}(m^2N)$ algorithm in itself.
In particular, the stabilized Gram-Schmidt algorithm requires $\sim 2 m^2N$ operations \cite{Golub}.

Backward evolution of the CLVs coefficients via Eq.~(\ref{clv-dyn}) does not require any explicit matrix 
inversion, but can be efficiently computed by back-substitution algorithms in $\sim m^3/3$ operations, 
while the normalization (\ref{norm-clv}) requires ${\cal O}(m^2)$ operations. Finally, $\sim m^2 N/2$
operations are needed in order to express the CLVs in the phase space coordinate basis via
Eq.~(\ref{clv-exp}).

Since the convergence of GS vectors and CLVs is exponential (see Section \ref{sec5} below), and in 
typical applications in systems with many degrees of freedom (where computational time may become an issue) one is interested to sample the geometrical structure of tangent space over the entire ergodic attractor, 
one usually has $t\gg 1$ and $k_{\rm dyn}\ll m$. For easy dynamics (which nevertheless covers the large 
majority of applications) this implies that the forward evolution is dominated by the QR decomposition, 
while the backward dynamics by the back substitution, for a total computational time 
\be
T_{\rm tot} \approx t\,m^2 \left(\frac{m}{3}  +\frac{N}{2} + 2 N\right) =t\,m^2 \left(\frac{m}{3} + \frac{5}{2} N\right) \,.
\ee 
Note that even when the full set of CLVs is computed, $m=N$, the
backward evolution is approximately 2.4 times faster than the forward
one, or 6 times if one is not interested in expressing the CLVs in the phase space coordinate basis. 
The situation is slightly less favorable for hard dynamics, which can be dominated by the forward 
evolution in tangent space, requiring full matrix multiplications every time step, for a total 
time $T_{\rm tot} \sim t\,k\, m\,N^2 $.

\item
{\it Transient length} --
As mentioned in the previous subsection, the convergence rates in both the
forward and backward transient rates
are exponential and related to the difference between consecutive
Lyapunov exponents (see also Section \ref{sec5} below). For instance, the convergence to the $i$-th CLV ($i>1$)
depends on the difference $\lambda_i-\lambda_{i-1}$. Dynamical systems with many degrees of freedom 
are typically characterized in the $N \to \infty$, by a limit spectrum spectrum $\lambda(i/N)$ that 
is piecewise continuous \cite{limitS} (possibly after removing those exponents connected to collective 
and sub extensive modes \cite{Takeuchi2009,Takeuchi2011}). 
This implies that the difference between consecutive exponents scales
to zero as $1/N$, so that it is advisable, when performing a finite
size analysis of such systems, to scale the transient time with the
number of degrees of freedom, $t_0 \to t_o N$.
\end{itemize}

\subsection{Non-invertible dynamics}
We finally discuss the extension of the concept of CLVs and of the dynamical algorithm to non-invertible 
dynamics. Mathematically, the construction of Oseledets' splitting requires the uniquely determined 
future and past images of any phase space point ${\bf x}_n$ on the attractor.
Although it is impossible to meet such a requirement in a non-invertible dynamical system (such as, e.g.
a chain of logistic maps), the difficulty may be circumvented by identifying the past images as 
{\it those ones that have been actually visited during the forward evolution} leading to ${\bf x}_n$. 
In this way, we can ``artificially'' restore the uniqueness of past images, and be free to apply 
all the machinery described above for invertible dynamics. 

There is of course a price to pay: some points ${\bf x}_n$ on the ergodic attractor can be reached by
following different past trajectories. This means that CLVs are not uniquely defined, but may depend
on the past trajectory followed to reach a given point. There is no longer a unique CLV 
in each point in phase space. This is not really an obstacle whenever one, instead of being
intersted in reconstructing the local tangent-space structure, aims at determining general statistical
properties of the CLVs.

\section{Convergence towards the covariant Lyapunov vectors}
\label{sec5}

We now analyze more closely the convergence of the dynamical algorithm
towards the CLVs, making use of fluctuations of finite time Lyapunov exponents.
Let us assume that the forward evolution has been carried out for a long-enough time to ensure a proper 
convergence to the maximally expanding subspaces (i.e. the GS vectors) along a given trajectory 
${\bf f}^{(n)}({\bf x}_0)$ for $n=0,1,\ldots,t+t_0$, and that the CLVs ${\bf v}^{(i)}_n$ are also
known at each point ${x}_n$ for $n=0,1,\ldots,t$. 
If $t_0\gg1$ we can assume these CLVs to be perfectly converged.

Now repeat the backward procedure starting for some randomly chosen vector at time $n+k \ll t+t_0-n$,
\be
\label{clv-exp2}
{\bf u}^{(i)}_{n+k} = \sum_{j=1}^i \tilde{c}_{n+k}^{(j,i)} {\bf g}^{(j)}_{n+k} \;
\ee
and evolve backward up to time $n$.
This procedure, allows to study the convergence towards ${\bf v}^{(i)}_n$ of the ``approximate'' CLV ${\bf u}^{(i)}_n(k)$ 
as a function of the finite backward evolution time $k$ on which the vector depends.
Let us define $\delta u^{(i)}_n(k)= |{\bf u}^{(i)}_n(k)- {\bf v}^{(i)}_n|\,$ which, being both vectors 
with unit norm, is just a function of their relative angle $\theta^{(i)}_n(k)$, 
\be
\delta u^{(i)}_n(k) = \sqrt{2} \, \sqrt{1-\cos \theta^{(i)}_n(k)}
\ee
(note that $\delta u^{(i)}_n(k) \sim \theta^{(i)}_n(k)$ for small angles).

Since we are interested in the average convergence rate towards the CLVs, we 
consider the generalized distance 
\begin{equation}
\Delta(k,q) = \langle [\delta u^{(i)}_n(k)]^q\rangle^{1/q}
\label{q-dist}
\end{equation} 
where $\langle \cdot \rangle$ denotes an average over both the ending points ${\bf x}_n$ 
and the different initial conditions $\tilde{c}_{n+k}^{(j,i)}$ in tangent space. Moreover, in order to
lighten the notations, we have dropped the dependence on the index $i$ (that is kept only
when strictly necessary).

\begin{figure}[tbp]
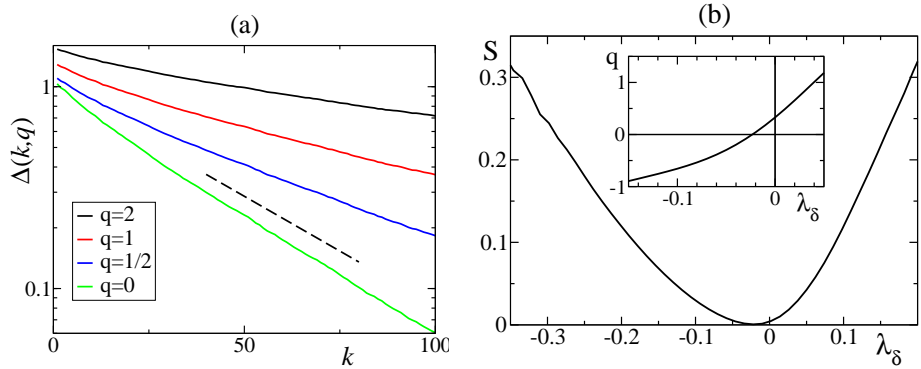

\begin{center}
\includegraphics[width=0.45\columnwidth]{conv_hen}
\hspace{0.1 cm}
\includegraphics[width=0.45\columnwidth]{multi_hen}
\vspace{0.05 cm}
\caption{(color online) Convergence of the second CLV in a chain of 5 coupled H\'enon maps 
(periodic boundary conditions) as defined in Eq.~(\ref{eq:henchain}) for $a=1.4$, 
$b=0.3$ and $\varepsilon = 0.025$. The uncertainty $\Delta(k,q)$, defined in 
Eq.~(\ref{q-dist}), is plotted in panel a for different
$q$-values. The dashed black line marks exponential decay with a rate $\lambda_2-\lambda_1$.
In panel b, the corresponding
large deviation function is plotted (by following the procedure discussed in 
the text). Finally, the inset contains the corresponding curve $q(\lambda_\delta)$.}
\label{fig:hen}
\end{center}
\end{figure}

Let us now consider a chain of H\'enon maps \cite{torcini92},
\begin{equation}
\hspace{-1.5cm} x_{n+1}(\ell)=a- \left[ x_n(\ell) + 
\varepsilon \big( x_{n}(\ell-1) - 2 x_n(\ell) + x_n(\ell+1)\big)
\right]^2 +  bx_{n-1}(\ell)  \, ,
\label{eq:henchain}
\end{equation}
where $i=1,\ldots,N$ 
is a lattice index and periodic boundary conditions are assumed. 
We have studied a set of $N=5$ maps for $a=1.4$, $b=0.3$, and $\varepsilon=0.025$.
For such parameter values there are 5 positive Lyapunov exponents and 5 (quite
large) negative ones. We focus on the the convergence towards the
second CLV following the procedure sketched above (the first one is automatically
generated during the forward iteration).\\
In Fig.~\ref{fig:hen}a we plot the behaviour of the generalized
distance (\ref{q-dist}) as a function of the backward convergence time
for four different $q$ values.
We see that the different $q$-distances all converge to $0$ exponentially, thus confirming  
the quick convergence of the method. 
However, different $q$-distances converge with different exponential rates.\\
In order to describe the process, it is reasonable to conjecture 
that the evolution towards the $i$-th CLV, as given by Eqs.~(\ref{components}-\ref{clv-conv}),
is ultimately controlled by the largest ratio between the various subspaces growth factors 
$\gamma^{(i)}_{k,n}$.  This is tantamount to assuming that all other off-diagonal 
terms have already converged to zero and at some time $m<n+k$ the vector ${\bf u}_{n'}^{(i)}(k)$ 
lies in the plane defined by the $(i-1)$-th and the $i$-th CLVs.

In this plane, the dynamics is controlled by the fluctuations of the finite time difference of the two FTLE,
$\lambda_\delta = \lambda_i(k) - \lambda_{i-1}(k)$ (we implicitly assume that 
the spectrum is not degenerate in $i$), 
which for large $k$ converges to the difference between the two corresponding 
Lyapunov exponents, $\Delta \lambda = \lambda_i - \lambda_{i-1}<0$. 
Therefore, for large times, ${\bf u}_m$ tend to rotate towards the $i$-th CLV and their relative angle
$\theta_m(k)$ approaches zero for a large majority of initial conditions.

We now consider the restricted generalized distance 
\begin{equation}
\Theta^q(k,q) \simeq \langle \theta_m(k)^q \rangle
\end{equation}
where the average has been restricted to small angles only. 
For small angles we may also write
\be
\theta_m(k) \simeq \theta_0 {\rm e}^{\lambda_\delta k}\,.
\ee
Accordingly, the fluctuations of $\lambda_\delta$ can be described by a large-deviation
function \cite{LD} $S(\lambda_\delta)$, 
\begin{equation}
P(\lambda_\delta,k) \approx {\rm e}^{-S(\lambda_\delta )k}
\end{equation}
with $S(\Delta\lambda)=0$ being the minimum of $S$.
Thereby,
\begin{equation}
\hspace{-1.cm} \Theta^q(k,q) = \int_{-\infty}^0 d \lambda_\delta \,P(\lambda_\delta,k)\, \theta_m(k)^q 
= \int_{-\infty}^0 d \lambda_\delta\, {\rm e}^{k [\lambda_\delta q  - S(\lambda_\delta) ]} \equiv
  {\rm e}^{L(q)k}  
\label{eq1}
\end{equation}
where, the upper bound to the integration interval is needed to avoid the unphysical divergence of the angle.
As long as the maximum of the exponential integrated in Eq.~(\ref{eq1}) is reached for a 
negative $\bar{\lambda}_\delta$ (which is itself a function of $q$),
that is, if the maximum lies inside the integration interval, 
$L(q)/q$ coincides with the generalized Lyapunov exponent (difference)
$\Lambda_q$ \cite{Generalized}, and
\begin{equation}
L(q) = \Lambda_q \equiv \bar{\lambda}_\delta q - S(\bar{\lambda}_\delta) \quad,\;\;  q = S'(\bar{\lambda}_\delta)\,.
\end{equation}
It is easy to check that for $q\to0$, $\Lambda_q/q$ converges to the Lyapunov exponents difference 
$\Delta \lambda$ (which is the extremal point where $S(\lambda_\delta=\Delta\lambda)=0$).

\begin{figure}[tbp]
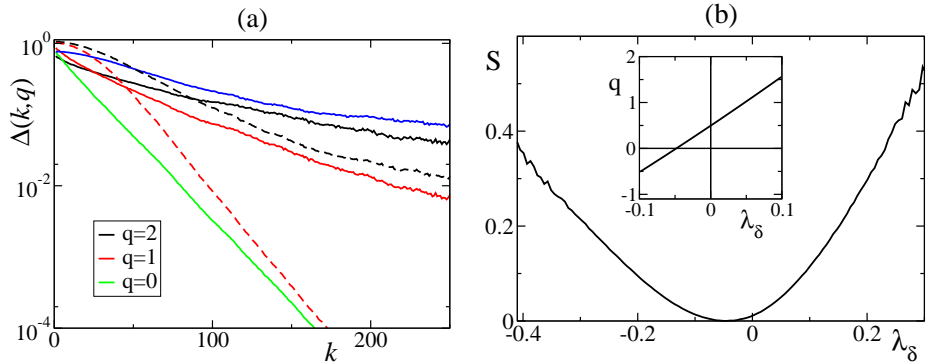

\begin{center}
\includegraphics[width=0.45\columnwidth]{conv_fpu}
\hspace{0.2 cm}
\includegraphics[width=0.45\columnwidth]{multi_fpu}
\vspace{0.05 cm}
\caption{(color online) Convergence of the 3rd and 9th CLVs in a chain of 5 FPU
  oscillators and energy density $e=5$. (a) Uncertainty $\Delta(k,q)$
  for different $q$-values and both the 3rd (solid lines) and 9th
  (dashed line) vectors. (b) Large deviation function $S$ for the 3rd
  CLV. In the inset the corresponding curve $q(\lambda_{\delta})$ is shown.
Additional technical details are
mentioned in the text.}
\label{fig:fpu}
\end{center}
\end{figure}

On the other hand, if $S(\lambda_\delta)$ extends to the positive semi-axis,
there always exists a critical 
\be
q_c = S'(0)
\label{eq:qc}
\ee
above which, the maximum is attained for $\lambda_\delta=0$ and 
\begin{equation}
L(q) = - S(0) \,.
\end{equation}
To resume, we have that $L(q)$ is non positive and monotonically
decreasing for $0\le q \le q_c$,
from $L(0)=0$ to $L(q_c)=-S(0)$. For $q>q_c$, finally, we have $L(q)=-S(0)$.\\
However, this is not yet the end of the story, since so far we have restricted our generalized distance to 
small angles. In fact, there also exists a family of special initial
conditions $\theta_0$ that are still close to the $(i-1)$-th CLV -- i.e. the ``wrong" 
direction -- at time $k$. 
They are those for which
$\theta_0$ is extremely close to $\pi/2$. The fraction of such trajectories is $\exp(\lambda_\delta k)$,
if $\lambda_\delta<0$, and is of order 1 otherwise. Altogether, their contribution $\bar{\Theta}^q(k,q)$
to the total generalized distance is
\begin{equation}
\bar{\Theta}^q(k,q) = \int_{-\infty}^0 d \lambda_\delta \,{\rm e}^{k [\lambda_\delta  - S(\lambda_\delta) ]} =
  {\rm e}^{L(1)k}  \,.
\end{equation}
We are now in a position to evaluate the two contributions combined,
\be
\Delta_q(k)= \left[\Theta^q(k,q) + \bar{\Theta}^q(k,q)\right]^{1/q}\,.
\ee
If $q_c<1$ then $L(1)=L(q_c)$ and the contribution from $\theta_0
\simeq \pi/2$ never prevails.\\
If $q_c>1$, on the other hand, the large angles contributions prevails
for $q \ge 1$. Note also that if the large deviations are restricted
to negative values, $q_c$ does not exist (being formally infinite)
and this latter case applies. 
To sum up, the generalized distance decays exponentially as
\be
\Delta(k,q) \sim \left\{
\begin{array}{lr}
{\rm e}^{k\,L(1)/q} & \mathrm{for}\;\;q\ge {\rm min}(q_c, 1)\\
{\rm e}^{k\,\Lambda_q/q} & \mathrm{for}\;\;q<{\rm min}(q_c, 1)\\  
\end{array}\right.
\ee
where $q_c$ is given by Eq.~(\ref{eq:qc}). Furthermore, for $q \to 0$, the
exponential convergence rate to the $i-th$ CLV ($i>1$) is given by the Lyapunov exponent
difference $\Delta\lambda_i = \lambda_i - \lambda_{i-1}$, while it
decreases monotonically as $q$ is increased.\\
In Fig.~\ref{fig:hen}b we plot the large deviation function for $\lambda_\delta = \lambda_2-\lambda_1$
for the above mentioned chain of H\'enon maps.
In order to minimize finite-size effects, $S(\lambda_\delta)$ has been estimated by comparing 
the histograms for two different time lengths $k_1<k_2$, 
\begin{equation}
S(\lambda_\delta) = \frac{\log P(\lambda_\delta,k_2) -\log P(\lambda_\delta,k_1)}{k_2-k_1}\,.
\end{equation}
The result for $k_1=30$ and $k_2=40$ is plotted in Fig.~\ref{fig:hen}b. In the inset we plot the derivative
of $S$, which coincides with $q$. The crossing of $q(\lambda_\delta)$ with the horizontal axis identifies the
minimum of $S$, i.e. the long time average of of $\lambda_\delta$ that turns out to be rather consistent with the
direct estimate of $\Delta \lambda = \lambda_2-\lambda_1$ ($-0.023$ instead of $-0.025$). The crossing with the vertical axis
identifies $q_c$, which, in this case is $q_c=0.3$, so that we are in the case $q_c<1$.\\
We are now in the position to compare the direct results plotted in Fig.~\ref{fig:hen}a with the
prediction of the multifractal analysis. First of all we note that the exponential decay for $q=0$ is in perfect agreement
with the Lyapunov exponents difference $\Delta \lambda$ (see the
dashed line). 
As for the other $q$-distance, since they are all
larger than $q_c$ their exponential decay rates are expected to be equal to $S(0)/q$. By computing the exponential 
rate for the largest times, we find that they are consistent with the
predicted $1/q$ behavior, although with an estimated prefactor
value of $S(0) \approx 0.008$. On the other hand a direct estimate from Fig.~\ref{fig:hen}b
yields $S(0) = 0.004$: the prefactor difference is relatively large, but definitely compatible with 
the amplitude of the finite-size corrections. Notice that since the minimum of $S(\lambda_\delta)$ occurs
for a small $\lambda_\delta$ value, $S(0)$ is quadratically small and this implies that a direct estimate
is rather problematic.\\
To test our prediction in a different system, we have also studied a
Hamiltonian model, namely a chain of Fermi-Pasta-Ulam (FPU)
oscillators \cite{FPU}, again with periodic
boundary conditions,
\begin{equation}
\ddot q_i = F(q_{i+1}-q_i)-F(q_i-q_{i-1})
\label{eq:FPU}
\end{equation}
where $F(x) = x - x^3$. In this model the energy $H = \sum \dot q_i^2/2 + (q_{i+1}-q_i)^2/2 +
(q_{i+1}-q_i)^4/4$ is conserved. We have studied a chain with $N=5$ oscillators (and thus $10$ 
degrees of freedom) and energy density $e=H/N=5$.
The convergence of the 3rd and 9th covariant vectors are reported in Fig.~\ref{fig:fpu}a (see solid and
dashed lines respectively).
The $0$-distance convergence of the two vectors is controlled by the differences $\lambda_3-\lambda_2$ and 
$\lambda_9-\lambda_8$, respectively. Given the symmetry of the Lyapunov spectrum 
($\lambda_i = \lambda_{N+1-i}$), the two vectors are expected to converge asymptotically 
in the same way, and this is a another way of checking the correctness of our approach.
In fact, we see that the three pairs of curves corresponding to different $q$-values values 
become eventually parallel. The initial difference is due to the fact that the 9th vector 
actually evolves in a higher dimensional space, where additional (and faster) relaxation 
processes are present.\\
Furthermore, we have studied the large deviation function for the difference between the
3rd and 2nd Lyapunov exponent (by comparing finite-time Lyapunov exponents for a time 
$t_1=10$ and $t_2=20$.  The results are plotted in Fig.~\ref{fig:fpu}b.
There we see that the minimum of $S(\lambda_\delta)$ occurs for $\lambda_\delta = -0.05$ 
that is quite close to the asymptotic value $\Delta \lambda= -0.053$. Moreover, we have $q_c \approx 0.5$, 
i.e. this case is borderline between the two above mentioned classes.
Finally, the ``asymptotic" 
exponential rates for the curves with $q=1$ and $q=2$ in Fig.~\ref{fig:fpu}a are 
$-0.066$ and 0.014, hinting at $S(\lambda_\delta)\approx 0.013$, a value close to the direct 
observation $0.011$ (from the inset in Fig.~\ref{fig:fpu}b), thus
confirming our theoretical argument.\\
To summarize, we have shown that our algorithm converges exponentially
to the true CLVs. However, when decay rates are computed via finite
time ensemble averages, only the $0$-norm decay rate coincides with
the difference between two consecutive Lyapunov exponents. Higher norm
decay with smaller exponential rates due to fluctuations of the FTLEs.
The norm-dependence can be explained in the context of large deviation theory.

\section{Static algorithms for covariant Lyapunov vectors}
\label{Sec6}
As anticipated, there exist various algorithms for the computation of CLVs. 
An alternative to the dynamical approach described in Sec.~\ref{Sec4} is represented by ``static'' 
algorithms, which do not make use of the intrinsic stability when the backward evolution is
restricted to suitable subspaced, but rather, determine the CLVs as linear combinations of 
either forward or backward GS vectors. More precisely, at each point ${\bf x}$ along a
given trajectory one has (for the sake of simplicity, in this section we omit the time index)
\be
{\bf v}^{(i)}=\sum_{j=1}^i \langle {\bf g}^{(j)}_+ | {\bf
  v}^{(i)}\rangle {\bf g}^{(j)}_+ =\sum_{j=i}^N \langle {\bf g}^{(j)}_- | {\bf v}^{(i)}
\rangle {\bf g}^{(j)}_-
\label{WS1}
\ee
where ${\bf g}^{(j)}_+$ and ${\bf g}^{(j)}_-$ are, respectively, the
forward and backword asymptotic GS vectors.

\subsection{Wolfe-Samelson algorithm}
A first static algorithm was introduced by Wolfe and Samelson in Ref.~\cite{WS}. \\
From Eq.~(\ref{WS1}), by using of the identity relation
\be
\sum_{k=1}^N \langle {\bf g}^{(i)}_+ | {\bf g}^{(k)}_-\rangle \langle
{\bf g}^{(k)}_- | {\bf g}^{(j)}_+\rangle = \delta_{i,j}
\label{WS2}
\ee
and with the help of simple algebraic manipulations, one obtains
(see \cite{WS} for more details)
\be
\sum_{j=1}^h\sum_{k=i}^{h-1} \langle {\bf g}^{(i)}_+ | {\bf g}^{(k)}_-\rangle 
\langle {\bf g}^{(k)}_- | {\bf g}^{(j)}_+\rangle c^{(j,h)}_+=0\;\;\;\;i\leq h
\label{WS3}
\ee
where $c^{(j,h)}_+ = \langle {\bf g}^{(j)}_+ | {\bf v}^{(h)}\rangle $
is the CLV coefficient expansion on the forward GS basis.\\
Eq.~(\ref{WS2}) can be recast in a matrix form, introducing the
$h\times h$ square matrix 
\be
[{\bf Q}^{(h)}]_{ij}=\sum_{k=i}^{h-1} \langle {\bf g}^{(i)}_+ | {\bf g}^{(k)}_-\rangle 
\langle {\bf g}^{(k)}_- | {\bf g}^{(j)}_+\rangle \;\;\;\;i,j\leq h
\label{WS4}
\ee
and the vector 
\be
[{\bf y}^{(h)}]_k = \langle {\bf g}^{(k)}_+ | {\bf v}^{(h)}\rangle\;\;\;k\leq h \, ,
\ee
namely,
\be
{\bf Q}^{(h)}{\bf y}^{(h)}=0\,.
\label{WS5}
\ee
Eqs.~(\ref{WS3}) and (\ref{WS5}) hold for $h>1$ (the first CLV
trivially coincides with the first forward GS vector) and allow 
obtaining the expansion coefficients of the $h$-th CLV as the kernel of a
matrix computed from the first $h$ forward and $h-1$ backward
asymptotic GS vectors
(the latter being the stable ones for the time-reversed dynamics).

\subsection{Kuptsov-Parlitz algorithm}
More recently, Kuptsov and Parlitz \cite{KP} have introduced a 
similar procedure, which makes use of LU factorization.\\
By using the matrix notation introduced in Section \ref{Sec4}, Eq.~(\ref{WS1}) 
can be rewritten as
\be
{\bf V}={\bf G}_+ {\bf C}_+= {\bf G}_-{\bf C}_-
\label{KP1}
\ee
with the plus and minus indices referring to forward and backward 
dynamics, respectively.
Since the forward matrix ${\bf C}_+$ is upper triangular, while
the backward one ${\bf C}_-$ is lower tringular, Eq.~(\ref{KP1}) can be
easily recast as an LU factorization \cite{Golub}
\be
{\bf P}{\bf C}_+= {\bf C}_-
\label{KP2}
\ee
where
\be
{\bf P}={\bf G}_-^T {\bf G}_+
\ee
Once again, if we are only interested in the $j$-th CLV, that is,
in the $j$-th column of the matrix ${\bf C}_+$, only
the $(h-1)\times h$ upper left corner of matrix ${\bf P}$ is needed. 
Since ${\bf C}_-$ is lower triangular and the first $j$ entries of its
$j$-th column are all zeros, we are left with the following system
of $(j-1)$ linear homogeneous equations in $j$ variables
\be
\sum_{i=1}^j [{\bf P}]_{k,i} \,[{\bf C}_+]_{i,j}=0\;\;k=1,2,\ldots,j-1
\ee
which defines the vector $j$ up to a rescaling factor. 
Not surprisingly, one needs the first $j$ forward and $(j-1)$ backward
asymptotic GS vectors to obtain the upper left part of ${\bf P}$
which is needed to compute the (forward) expansion coefficients of the
first $j$ CLVs.

\subsection{Comments}

To conclude, all static methods, be the direct intersection method of
Eqs.~(\ref{ome}-\ref{ose2}) or the two refinements briefly discussed
above, require the solutions of certain systems of linear homogeneous
equations, which in turn depend on the forward and backward GS
vectors. For direct subspace intersection, the first $j$ forward and the last
$(N-j+1)$ backward GS vectors are required to compute the first $j$ CLV, 
while only the first $j$ forward and $(j-1)$ backward vectors are needed
by the refined algorithms - an improvement if one is interested in
either the first or last CLVs. 

However, in our opinion static approaches suffer from a number of
disadvantages. First of all, one has to compute both forward and
backward GS vectors, and thus is forced to perform twice
vector orthonormalization at each trajectory point where CLVs are
needed. Note that, as discussed in Section \ref{Sec4B}, vector
orthonormalization is the computationally most demanding part of the dynamics for both short 
ranged and globally coupled large dynamical systems. Therefore, it is
not a good idea to double the number of such orthonormalizations with
respect to the plain dynamical algorithm of Section \ref{Sec4}.\\
A second concern regards the solution of large systems of linear
equations, which has to be performed by singular value decomposition
(SVD) to attain a satisfactory numerical accuracy. 
However, SVD is more time consuming than back substitution by a factor
18, as it requires $\sim 6 m^3$ operations for an $m\times m$ matrix \cite{Golub}.

Finally, for what regards memory requirements, only forward GS vectors need to be stored 
by the static algorithm. While this reduces the memory requirement to
about $2/3$ of what needed by the dynamical algorithm to store both
${\bf G}$ and ${\bf R}$ matrices, this memory advantage is lost
whenever one is only interested in the angles between vectors, for which
the dynamical algorithm only needs to store the upper triangular
matrices ${\bf R}$ and thus finds itself in a better position.

\section{Angles between covariant vectors and subspaces}
\label{sec7}
Covariant Lyapunov vectors provide direct information on the
geometrical structure of tangent space. In particular, angles and
(near)-tangencies between different CLVs or their associated subspaces can be
used to characterize the dynamical properties of a chaotic dynamical
system. In most applications of interest, this amounts to studying the
distribution of such angles over the (ergodic) attractors. 
While angles are not invariant under a generic coordinate
transformation, non-singular transformations do preserve zero and
non-zero angles. Note also that CLVs corresponding to non-degenerate LEs 
may not become completely parallel along a given trajectory. Since
they evolve via Eq.~(\ref{clv1}), it is clear that should at any point
two vectors be completely parallel, they would stay the same along the
entire trajectory, contradicting the non-degeneracy
assumption. However, trajectories can pass arbitrarily close to such
tangent points, resulting arbitrarily small angles. Therefore, relevant physical information has to be
encoded in the way the angles probability distributions approaches the
null angle. 

In Refs.~\cite{Takeuchi1}, angles between CLVs have been
studied to show that the tangent space of generic spatially
extended dissipative systems, such as the Kuramoto-Sivashinsky equation
or the complex Ginzburg-Landau equation, is split into two
decoupled subspaces. One comprises a finite number of frequently
``entangled" CLVs, or {\it physical} modes, which carry all the relevant information of
the trajectory. A second residual set is composed of strongly decaying {\it spurious}
modes which are transversal to the ``physical'' manifold and 
themselves organized in mutually transversal subspaces. The number of physical
modes, which is extensive in the system total degrees of freedom $N$,
can be interpreted as the number of {\it effective} degrees
of freedom needed to faithfully describe the chaotic dynamics, leading to
the conjecture that the physical modes constitute a local
approximation of the inertial manifold \cite{YR_PRL}.\\
Another interesting issue regards the degree of non-hyperbolicity of a
dynamical system. In hyperbolic systems \cite{RuelleLL, PesinBook}, there exists a direct sum
decomposition of the tangent space $\mtT_x \mtM$ at each point $x$
into three {\it invariant} subspaces 
\begin{equation}
\mtT_x \mtM = 
{\bf E}_x^u \oplus {\bf E}_x^s \oplus {\bf E}_x^0\,. 
\end{equation}
The unstable subspace ${\bf E}_x^u$ is spanned by the
CLVs associated to positive LEs, so that each vector ${\bf u}\in {\bf
  E}_x^u$ is exponentially contracted backward in time. Similarly, the
stable subspace ${\bf E}_x^s$ is spanned by the CLVs associated to negative LEs and any ${\bf u}\in {\bf
  E}_x^s$ is exponentially contracted forward in time. Finally, ${\bf E}_x^0$ is associated to the null 
LEs and their corresponding
covariant subspace\footnote{It is non-empty only for continuous time flows and/or
  in the presence of conservation laws which reduce the attractor
  dymensionality.}. 
In particular, hyperbolicity implies that the stable and
unstable subspaces are nowhere tangent.
Dynamics on a hyperbolic attractor is structurally
stable, i.e., is insensitive to variations of parameters. It manifests
strong stochastic properties which allow for detailed theoretical
analysis and useful results such as the shadowing lemma
\cite{bowen}. Violations of hyperbolicity may manisfest themselves either as a
change along the attractor of the number of stable and unstable
directions ({\it unstable dimension variability}) or through the
presence of {\it homoclinic tangencies}, i.e. points where the
stable and unstable manifolds are mutually tangent.

The fact that the overwhelming majority of dynamical systems of any practical use 
are not hyperbolic may seem a serious drawback, but the ``chaotic'' hypothesis \cite{chaohyp} assumes that
generic chaotic systems, in spite of violations of hyperbolicity, can
be treated as essentially hyperbolic ones, allowing one to extend a
number of useful results to non-hyperbolic chaotic systems. In
particular, it implies ergodicity of the attractor and the existence
of well-defined time averages with a probability distribution
satisfying a large deviation law.

While the chaotic hypothesis has an evident empirical success, as testified
by the positive tests of the fluctuation theorem \cite{Lepri, Ciliberto}, it is
nevertheless of great interest to assess the degree of violation of
hyperbolicity in a given chaotic system, especially in relation to its
dynamical properties.
In the following, we
concentrate on homoclinic tangencies. 

\subsection{Subspace intersection}
As it has been correctly pointed out in Ref. \cite{KK}, in order to
compute the angle between two linear subspaces it is not sufficient to
compute all the angles between pairs of vectors taken from two
bases spanning the two subspaces. Indeed one has to consider angles 
between arbitrary linear combinations of such vectors. As shown
in \cite{KK}, this can be taken into account by SVD. Consider two generic
subspaces, respectively spanned by $m_1$ and
$m_2$ different CLVs, $m_1+m_2\leq m$, with $m$ being the tangent space
dimension. Suppose $m_2 \geq m_1$. First of all, we organize the CLVs from the two subspaces in two
different matrices: ${\bf U}_1$ (of size $m\times m_1$), whose columns contain the vectors
from $V_1$ and ${\bf U}_2$ ($m\times m_1$), with the vectors
from $V_2$. Since angles can be equally computed in the phase space
coordinate basis as well as in the GS basis, we choose the latter and 
consider the proper columns of the matrix ${\bf C}$. 

We have to compute the QR factorization of both matrices,
\be
{\bf U}_1={\bf Q}_1 \,{\bf R}_1\;\;\;,\;\;{\bf U}_2={\bf Q}_2 \,{\bf R}_2
\ee
and compose the $m_1 \times m_2$ matrix 
${\bf Q}_2^T \,{\bf Q}_1$.
There are $m_1$ principal angles $\theta^{(i)}_n \in [0,\pi/2]$ between the two subspaces, and their
cosines are given by the principal values $s^{(i)}$ of ${\bf Q}_1^T \,{\bf
  Q}_2$, that is
\be
s^{(i)}_n=\cos \theta^{(i)}_n\;\;\;i=1,2,\ldots,m_1
\ee
where we have explicated the time index $n$. 

In our case we are interested in the intersection between the stable
${\bf E}^s$ and the unstable ${\bf E}^u$ manifolds. 
Note that since the unstable manifold ${\bf E}^u$ is spanned by the
first $m_1$ CLVs, the matrix ${\bf U}_1$ is upper triangular, and its
corresponding orthogonal matrix ${\bf Q}_1$ is just the identity
matrix, thus simplifying the calculations of principal angles.
In particular, we are interested in the minimum angle
$\bar{\theta}_n=\mathrm{min}_i \,\theta_n^{(i)}$.

\subsection{Numerical examples}

In this section, we present two numerical examples: a Hamiltonian system, the FPU chain
(\ref{eq:FPU}) with $N$ oscillators, and a dissipative one, a chain of
$N$ H\'enon maps (\ref{eq:henchain}). 
In the Hamiltonian
case we have $m_1=m_2=N-2$, since there are 4 null LEs in one dimensional FPU,
corresponding to momentum and energy conservation and the associated
symmetries. In the H\'enon chain case, one has typically $m_1=m_2=N$, and there
are no zero LEs.
We have investigated the tangent-space geometrical structure by
measuring the minimum angle between the stable and unstable manifolds
along a trajectory which samples the system's ergodic measure.

\begin{figure}[t]
\centering
\includegraphics[draft=false,clip=true,width=0.95\textwidth]{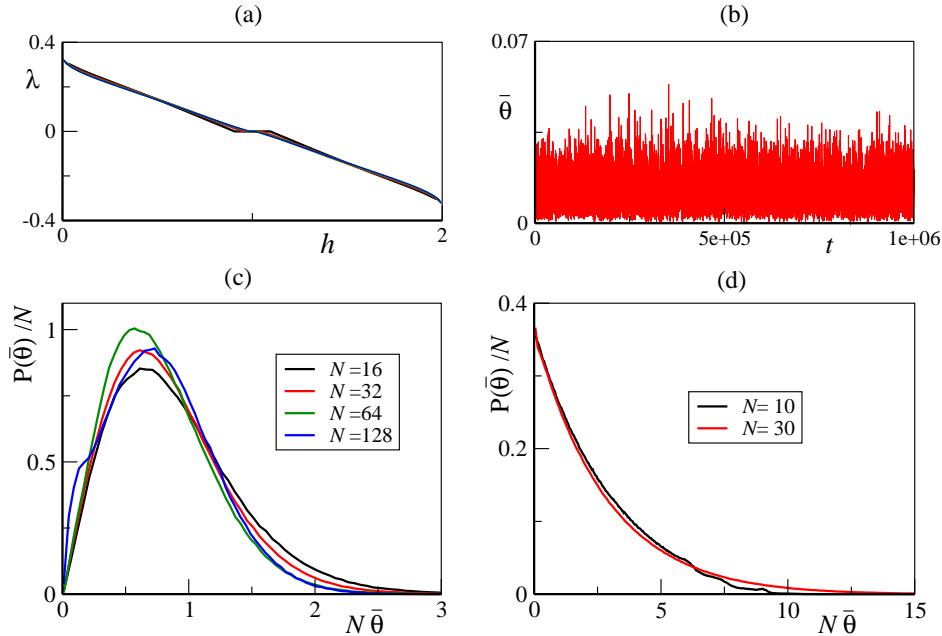}
\caption{(color online) FPU (at energy density $e=10$) and H\'enon chains results. (a) FPU
Lyapunov spectrum as a function of the
rescaled index $h=(i-0.5)/N$ for different system sizes,
$N=16,32,64,128$. (b) FPU
Minimum angle $\bar{\theta}$ timeseries for $N=64$. (c)
FPU probability distribution $P(\bar{\theta})$ as a function of the
rescaled minimum angle for
different system sizes, $N=16,32,64,128$. (d) Minimum angle
rescaled probability distribution for the H\'enon chain at sizes
$N=10$ and $N=30$. Other parameters as in Fig. \ref{fig:hen}} 
\label{figA1} 
\end{figure}

We start discussing the FPU system at an energy density $e=H/N=10$, which is
known to be characterized by a well developed chaotic dynamics.
Our numerical simulations employed periodic boundary conditions and a McLachlan-Atela integration
algorithm \cite{MA}, which is well-suited for Hamiltonian systems, and an
integration step of $\Delta t=0.05$ (we have verified the stability of
our results versus $\Delta t$). We employed a transient (both forward
and backward) of $t_0=N\cdot 10^5$ time units, and sampled the minimum
angle $\bar{\theta}_n$ between the stable and unstable manifold along
$10^6$ time units (our sampling and orthonormalization rate is
one time unit). 

We considered $e=10$ and different system sizes
between $N=16$ and $N=128$.
The rescaled Lyapunov spectra for this system are shown in Fig.~\ref{figA1}a, (they have been 
computed both along the forward dynamics with the Benettin {\it et al.} 
algorithm as well as from the CLVs expansion rates (\ref{clv1}), 
as a check of CLV numerical convergence). The stationary time-series 
corresponding to the minimum angle $\bar{\theta}_n$ between stable and 
unstable subspaces is plotted in Fig.~\ref{figA1}b, where one can see that
it is rather irregular. We find therefore convenient to reconstruct the 
probability distribution $P(\bar{\theta})$  (PDF) of the minimal angles.
(see Fig.~\ref{figA1}c). By comparing the distribution for different
system sizes, one can conclude that the minimum angle scales as
\begin{equation}
\bar{\theta}\sim \frac{1}{N}
\end{equation}
as testified by the relatively good data collapse of the various PDFs.
This is somehow intuitive, considered that the number of principal angles 
is equal to the stable and unstable manifold dimension. Notice that the 
probability distributions vanish algebrically (roughly linearly) as $\bar{\theta}
\to 0$, are characterized by a finite maximum and decay exponentially fast for large angles. 

This behavior can be compared with results for a chain of H\'enon
maps. Also for this dissipative case, the minimum angle scales as $1/N$, 
but the shape of the PDF is rather different: it starts with a finite value for 
$\bar{\theta} \to 0$ and decays esponentially as the angle is increased 
(see Fig.~\ref{figA1}d). Note also that these latter PDFs are
characterized by a much larger width.

One can certainly conclude that the dissipative system is characterized by
stronger violations of hyperbolicity. It is, nevertheless, important to remark
that a vanishing PDF when $\bar{\theta} \to 0$ does not imply hyperbolicity. In hyperbolic
systems, the distribution of minimum angles is {\it bounded away from zero}\cite{Ginelli2007}. 
A vanishing $P(\bar{\theta})$ implies that trajectories exist which pass arbitrarly close to a
(zero mesure) set of homoclinic tangencies\footnote{Here we assume CLVs in $\mtT_x \mtM$ are
continuous with respect to $x$. A sufficient condition for the continuity of the asymptotic GS base, 
which in turns implies the continuity of CLVs is discussed in \cite{EP}.}. 
The different behavior of $P(\bar{\theta})$ in zero reveals, however, important differences in 
the spatial structure of homoclinic tangencies. In particular, if $P(\bar{\theta})\sim
\bar{\theta}^\alpha$ for $\bar{\theta}<<1$, the measure of points characterized by angles 
(between the stable and unstable subspace) smaller than a certain threshold $\theta_0$ will scale as 
$\theta_0^{\alpha+1}$. The implications of this scaling law for the global dynamics are the 
object of current investigation.

\section{Concluding remarks}
\label{sec8}
Covariant Lyapunov vectors provide an intrinsic decomposition of tangent space that is 
invariant under time reversal and independent of the norm (CLVs coincide with the Floquet 
eigenvectors when computed along periodic orbits \cite{TrevisanPancotti}).
For this reason, whenever individual directions need be considered, CLVs
have to be chosen over other vectors, such as the asymptotic GS vectors or the 
``singular" vectors \cite{Legras}, which represent a finite-time version of the former ones,
used in ensemble forecast applications and which do depend on the norm. 
Indeed, it has been shown \cite{Ginelli2007} that the orthonormalization procedure 
(\ref{split1}) introduces, spurious structures in the individual GS vectors which
have nothing to do with the underlying dynamics.

It would be interesting to apply CLVs for a better control of the uncertainty in nonlinear 
models such as those used for weather forecast.  While variational data-assimilation techniques 
can benefit from the generic knowledge of the system's unstable manifold \cite{4var}, 
it is conjectured that the most relevant instabilities which affect large scale structures 
in atmospheric models are not those ones associated to the fastest time scales (i.e., the largest LEs), 
but rather to slower instability modes, characterized by longer spatial wavelengths. 
It logically follows that the most important directions for data-assimilation coincide with those
associated to small positive LEs, which can be individually accessed only by computing CLVs. 
One may therefore hope that the ability to access the single unstable directions associated to 
well-defined timescales (the inverse of the LE) is of practical use in optimal forecast methods. 

Covariant Lyapunov vectors have a number of further potential applications in dynamical
systems theory. They can be used to assess the hyperbolicity of the
underlying dynamics \cite{Ginelli2007,KK} and, as shown in the previous section, can also contribute to 
shed light on the spatial structure of tangent space. In this respect, it is worth mentioning
that preliminary simulations performed in the FPU chain at lower energies (below the
strong stochasticity threshold) reveal stronger violations of hyperbolicity.
Moreover, CLV have been also employed to characterize the collective dynamics of large 
chaotic systems through their localization properties \cite{Takeuchi2009}, thus allowing to
establish a connection between microscopic evolution and the emergence of global properties. 

Finally, one puzzling finding in Hamiltonian and symplectic systems regards the
$k^{-1}$ divergence displayed by the power spectrum of the spatial
part of the covariant vectors associated to the smallest LEs
\cite{Ginelli2007}. Further work is required to clarify whether this property
can be related to the so-called hydrodynamic
Lyapunov modes \cite{HLM} or whether this $1/f$ behavior has any relation
with actual dynamical properties of these systems.

\ack
We warmly thank G. Morriss, P. Kuptsov, S.P. Kuznetsov, A. Pikovsky and
M. Sala for enlightening discussions.\\

\end{document}